\documentclass[aps,pre,twocolumn,address,showpacs,amsmath,amssymb]{revtex4}

\usepackage[makeroom]{cancel}

\usepackage{graphicx}

\usepackage{verbatim}

\usepackage{float}

\usepackage{dcolumn}

\usepackage{bm}
\usepackage[normalem]{ulem}
\usepackage{color}

\usepackage{mhchem}

\usepackage{epstopdf}

\newcommand{\br}{\mathbf{r}}

\newcommand\underrel[2]{\mathrel{\mathop{#2}\limits_{#1}}}

\newcommand{\sep}{ \ \ \ , \ \ \ }

\newcommand{\beq}{\begin{equation}}
\newcommand{\eeq}{\end{equation}}
\newcommand{\beqn}{\begin{eqnarray}}
\newcommand{\eeqn}{\end{eqnarray}}

\newcommand{\pp}{\partial}
\newcommand{\dd}{{\rm d}}
\newcommand{\ee}{{\rm e}}

\newcommand{\eq}{Eq.\ }

\newcommand{\eqs}{Eqs.\ }
\newcommand{\fig}{Fig.\ }
\newcommand{\sect}{Sec.\ }

\newcommand{\cO}{{\cal O}}

\newcommand{\Pin}{P_{\rm in}}
\newcommand{\Po}{P_{\rm out}}
\newcommand{\Sin}{S_{\rm in}}
\newcommand{\So}{S_{\rm out}}

\newcommand{\Hin}{H_{\rm in}}
\newcommand{\Ho}{H_{\rm out}}
\newcommand{\HHin}{\mathcal{H}_{\rm in}}
\newcommand{\HHo}{\mathcal{H}_{\rm out}}
\newcommand{\fin}{f_{\rm in}}
\newcommand{\fo}{f_{\rm out}}

\newcommand{\foA}{f_{\rm out}^{(A)}}
\newcommand{\foB}{f_{\rm out}^{(B)}}

\newcommand{\foAp}{f_{\rm out}^{(A)'}}
\newcommand{\foBp}{f_{\rm out}^{(B)'}}

\newcommand{\foApp}{f_{\rm out}^{(A)''}}
\newcommand{\foBpp}{f_{\rm out}^{(B)''}}

\newcommand{\hatPin}{\hat{P}_{\rm in}}
\newcommand{\hatPo}{\hat{P}_{\rm out}}

\newcommand{\cred}{\color{red}}

\begin{document}

	\title{Chemical reaction-controlled phase separated drops:\\
		 Formation, size selection, and coarsening
	}
	
\author{Jean David Wurtz}
	\affiliation{Department of Bioengineering, Imperial College London, South Kensington Campus, London SW7 2AZ, U.K.}
\author{Chiu Fan Lee}
\email{c.lee@imperial.ac.uk}
\affiliation{Department of Bioengineering, Imperial College London, South Kensington Campus, London SW7 2AZ, U.K.}
	
	\begin{abstract}	
Phase separation under non-equilibrium conditions is  exploited by biological cells to organize their cytoplasm but remains poorly understood as a physical phenomenon. Here, we study a ternary fluid model in which  phase-separating molecules can be converted into soluble molecules, and vice versa, via chemical reactions. We elucidate using analytical and simulation methods how drop size, formation, and coarsening can be controlled by the chemical reaction rates, and categorize the qualitative behavior of the system into distinct regimes. Ostwald ripening arrest occurs above critical reaction rates, demonstrating that this transition belongs entirely to the non-equilibrium regime. Our model is a minimal representation of the cell cytoplasm.
	\end{abstract}
	\maketitle

Phase separation is a ubiquitous phenomenon in our physical world, ranging from cloud formation to oil drop formation in water 
 \cite{bray_advphys02}.
Recently, it is also realised that phase separation is exploited in the cell cytoplasmic organization in the formation of non-membrane bound organelles called ribonucleoprotein (RNP) granules \cite{brangwynne_softmatt11,hyman_annrev14}.
RNP granules are a diverse set of structures that play important roles in the functioning of the cell, from RNA processing and stress response \cite{anderson_currbiol09,protter_trends16}, to cell division \cite{zwicker_pnas14} and  germ line specification \cite{brangwynne_science09,voronina_coldspring11}.
However, the mechanisms enabling the rapid and controlled assembly and disassembly of RNP granules have only begun to be investigated. Chemical reactions, e.g., in the form of ATP-driven enzymatic reactions that convert one protein state to another (e.g., unphosphorylated to phosphorylated) are prime candidates for the cell to manifest controlled phase separation. 
For instance, such a scheme has been proposed as a mean to induce localised phase separation in the {\it C.~elegans.}~embryo \cite{lee_prl13,saha_cell16,weber_njp17}, and to organise the centrosomes prior to cell division \cite{zwicker_pnas14}.
However the physics of non-equilibrium phase separation driven by chemical reactions has only started to be investigated. 
For instance, non-equilibrium processes have been discussed in the context of lipid domains in plasma membranes \cite{Turner2005,Fan2008}.
More recently, it has been realised that 
although in equilibrium phase separation, a multi-drop, finite system will invariably coarsen to a single condensed drop via Ostwald ripening, chemical reactions can arrest this ripening process completely in a binary fluid 
\cite{glotzer_prl95,zwicker_pre15}. 
Here, we categorize  comprehensively and under general conditions, how  unimolecular reactions that convert a two-state molecule between a phase-separating state and { a} soluble state  can control drop formation, coarsening, and size selection. We achieve this by generalizing and improving upon the assumptions adopted in \cite{zwicker_pre15}. Specifically, contrary to \cite{zwicker_pre15}, we analyse the regimes of large drops and non negligible supersaturation, include the presence of cytosol by going beyond the binary fluid restriction, and allow for arbitrary equilibrium concentrations inside and outside drops.
Our model is arguably the minimal model relevant to the mechanism of chemical reaction-controlled phase separation in the cell cytoplasm.

\begin{figure}
	\centering
	\includegraphics{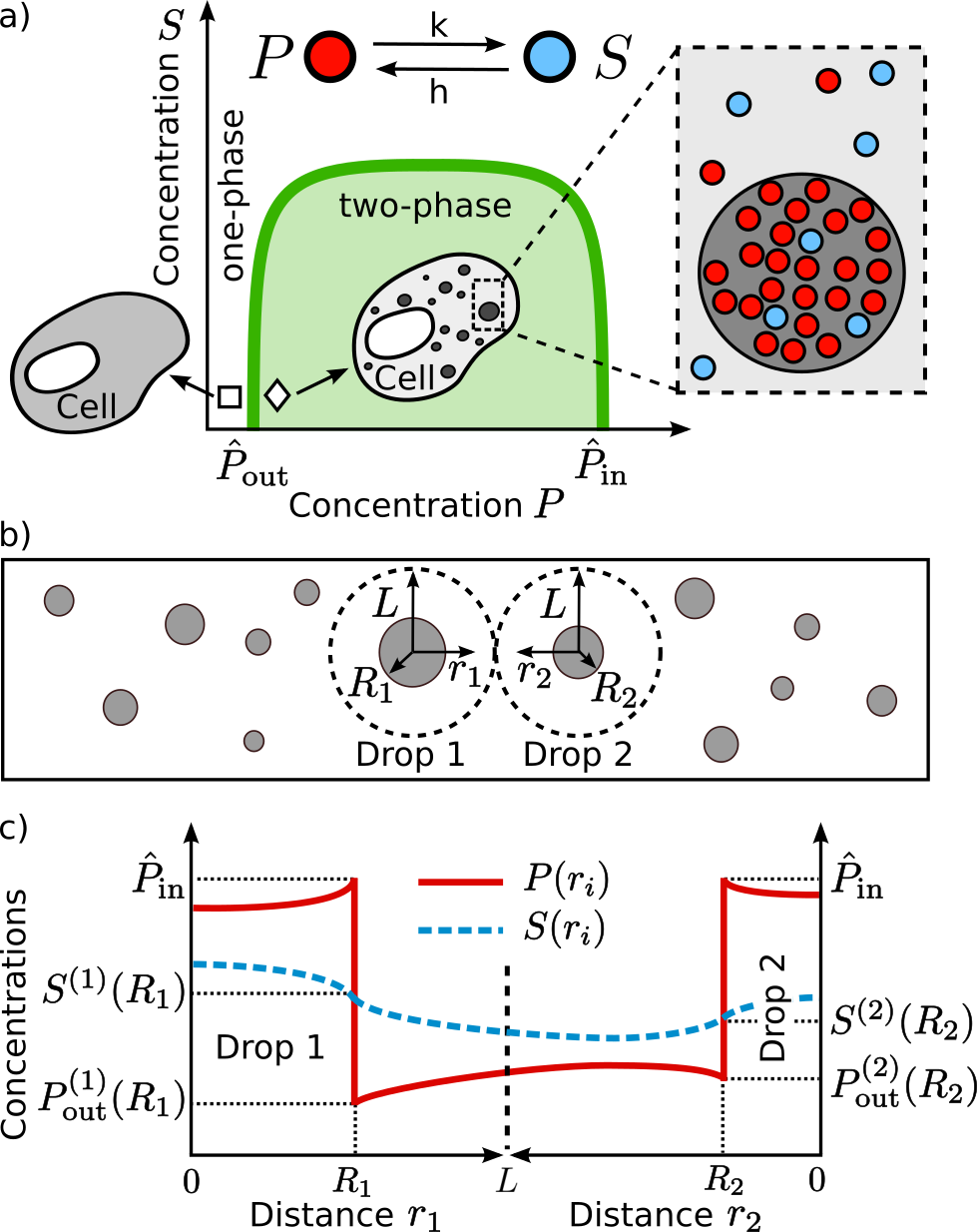}
	\caption{
		{\it Model of cytoplasmic phase separation.} 
		a) The cell cytoplasm is  modeled by a ternary fluid composed of phase-separating ($P$) and soluble ($S$) 
		{molecular states}, and other cytoplasmic components ($C$). Chemical reactions convert $P$ into $S$ at the rate $k$, and $S$ into $P$ at the rate $h$ (\eq \eqref{eq:PRL:reactions}). At equilibrium ($k=h=0$), the system is well mixed  (`$\square$') if the concentrations of $P$ and $S$ lie outside the phase boundary (green line in the phase diagram), and the system phase separates otherwise (`$\lozenge$').
In the latter case, we assume that $S$ does not phase separate and remains homogeneous.
		b) A multi-drop system with drop number density $\rho$ is studied by considering two interacting subsystems ($i=1,2$) of radius $L=[3/(4\pi\rho)]^{1/3}$, 
		each having a drop of radius $R_i$ in their center. 
		c) Schematics of the concentration profiles of $P$ and $S$ in the subsystems  when chemical reactions are present ($k,h>0$, \eqs \eqref{eq:PRL:profileIn},\eqref{eq:PRL:profileOut}). 
		At the subsystems' boundaries ($r_i=L$) the profiles and their derivatives are matched by assumption (\eq \eqref{eq:PRL:BC2}).
\label{fig:PRL:intro}
	}	
\end{figure}
	
Our ternary mixture consists of two molecular 
states, one phase-separating ($P$) and one soluble ($S$), plus the solvent or cytosol ($C$). States $P$ and $S$ can be converted into each other by the chemical reactions
\begin{equation}
\label{eq:PRL:reactions}
\ce{
$P$
<=>[$k$][$h$]
$S$
}
\end{equation}
where $k$ and $h$ are the reaction rate constants. The non-equilibrium nature of these reactions lies in the fact that both reaction rates are independent of the local concentrations and thus have to be driven by free energy consumption. 
In the context of the cell, these reactions can be, e.g., ATP-driven post-transcriptional protein modifications \cite{Alberts2008}, that affect protein phase-separating behavior. For example, the phase separation of intrinsically disordered proteins can be regulated via their phosphorylation/dephosphorylation \cite{li_nature12,bah_jbc16}.



At equilibrium ($k,h=0$), a finite system will inevitably coarsen via  Ostwald ripening \cite{lifshitz_jpcs61} and drop coalescence \cite{siggia_pra79}. Here, we assume that drop diffusion is negligible so we will focus exclusively on the Ostwald ripening. In the cell context, this is motivated by the
 strong suppression of macromolecular diffusion in the cell cytoplasm   \cite{weiss_biophysj04}.  Ostwald ripening results from two effects:
1) the Gibbs-Thomson relation dictating that for a drop of size $R$, the concentrations of solute 
{ inside and outside the drop next to the interface are}
$\hatPin$ and $\hatPo\left(1+l_c/R\right)$ respectively, where $l_c$ is the capillary length and $\hat P_{\rm in/out}$ are the equilibrium phase coexistence concentrations (see \fig \ref{fig:PRL:intro}a));
	and 2) the concentration profile of the solute in the dilute  phase is given by the  steady-state solution to the diffusion equation (the quasi-static assumption). 
  	These two effects combined lead to a diffusive flux of solute from small drops to big drops \cite{lifshitz_jpcs61}. 
 
When chemical reactions are switched on, we assume that local thermal equilibrium remains valid so that the interface boundary conditions for $P$ are unchanged \footnote{We have verified these conditions using simulation methods \cite{SI}.}. $S$ is considered inert to phase separation in the sense that its concentration profile is continuous across the interface \footnote{	
	This assumption  	
 	is not essential and we describe the more general case where $S$ is discontinuous at the drop interface in  \cite{SI}.}. In addition, we assume that the concentration profiles inside and outside the drops are given by: 	
\beqn
\label{eq:PRL:reacDiffP}
\frac{\pp P_{\rm in/out}}{\pp t} &=&D \nabla^2 P_{\rm in/out}-k P_{\rm in/out}+hS_{\rm in/out} 
\\
\label{eq:PRL:reacDiffS}
\frac{\pp S_{\rm in/out}}{\pp t} &=&D \nabla^2 S_{\rm in/out}+kP_{\rm in/out}-hS_{\rm in/out}
\ ,
\eeqn
where $P_{\rm in/out}$ and $S_{\rm in/out}$ denote the concentration profiles of $P$ and $S$ inside and outside drops with subscripts ``in'' and ``out'', respectively. For simplicity, we assume the same diffusion coefficient $D$ for both species and in both phases.

To see why Ostwald ripening can be arrested in our ternary mixture, we will now provide an intuitive argument based on a similar consideration for binary mixtures \cite{zwicker_pre15}. 
 We consider  a  homogeneous system of total solute concentration $\phi=P_{\rm tot} +S_{\rm tot}$ where $P_{\rm tot}$ $(S_{\rm tot})$ is the total concentration of $P$ $(S)$ in the system. If the supersaturation $\triangle = P_{\rm tot} - \hat{P}_{\rm out}$ is positive drops can be nucleated, initiating  phase separation (\fig \ref{fig:PRL:intro}a)). At small $\triangle$ 
 the drop density is low and drops only interact with the far-field  concentration. 
 We focus only on the early growth regime so that the supersaturation $\triangle$ remains close to $P_{\rm tot}-\hatPo$: for a drop of radius $R$, the diffusive profile leads to an influx of $P$ into the drop at the rate \cite{lifshitz_jpcs61}
 \beq
 \label{eq:PRL:influx}
4\pi D R\left( \triangle- \frac{\hatPo l_c}{R}\right)
 \ .
\eeq
At the same time, the chemical reactions inside the drop lead to a depletion of $P$ at the rate
\beq
\label{eq:PRL:degradationflux}
\frac{4\pi R^3}{3} k \hat{P}_{\rm in}
\ .
\eeq
As $R$  increases, the depletion rate will eventually surpass the influx from the medium, so that the balance between \eqs (\ref{eq:PRL:influx}) and (\ref{eq:PRL:degradationflux}) leads to a steady-state radius. In the limit of large $R$ so that we can ignore the term $\hatPo l_c/R$  (but still small such that  
	$\triangle \approx  P_{\rm tot} -\hatPo$), the steady-state $R$ is
\beq
\label{eq:PRL:Ru:intuitive}
\sqrt{\frac{3D \triangle}{k \hatPin} } \ .
\eeq
In other words, we expect that in a multi-drop system, the size of all drops are given by \eq (\ref{eq:PRL:Ru:intuitive}). We shall see that this regime in fact corresponds to the upper bound of stable $R$ in a multi-drop system (\fig \ref{fig:PRL:compare}).

In our argument above, we have neglected the reverse reaction $S\to P$ and the the effect of the chemical reactions on the diffusive profiles, which, as we shall see, can significantly change the  system's behavior. 
	We will now incorporate these effects into our analysis. We will also consider arbitrary supersaturations so that drops may be close to each other. As a result a far-field concentration may not exist, rendering the Lifshitz-Slyozov theory \cite{lifshitz_jpcs61} inapplicable.
Consider a multi-drop system such that drops are on average a distance 
	$2L$ apart
where $L$ is of the order  $\rho^{-1/3}$ with $\rho$ being the drop number density.   For simplicity, we will first focus on two spherical subsystems of radius $L$, each having a spherical drop in their center (\fig \ref{fig:PRL:intro}b)). We assume that the concentrations and their gradients at the  boundaries of the two subsystems match (\fig \ref{fig:PRL:intro}c)). 
The rational for this approximation is that in a multi-drop system, the actual boundary conditions  are influenced by many neighbouring drops and we treat these fluctuating boundary conditions in a {\it mean-field} manner by assuming spherical symmetry around the drops. In other words, the concentration outside a drop depends only on the distance from the drop centre. Moreover it is assumed that the two-drop system is stable (unstable) if the full multi-drop system is stable (unstable). The validity of this approximation will be verified later using Monte Carlo simulations.
The corresponding boundary conditions, besides the Gibbs-Thomson at the drops' interfaces, are
\beqn
\label{eq:PRL:BC2}
\quad \Po^{(1)}(L) = \Po^{(2)}(L), \quad \nabla_{\br_1} \Po^{(1)}|_{L} = - \nabla_{\br_2} \Po^{(2)} |_{L} \ ,
\eeqn
and the same apply to $S_{\rm in/out}^{(i)}$. The subscript $i=1,2$ denotes the drop index. Note that we use two different coordinate systems $r_1$ and $r_2$, each having their respective  drop's center as the origin (\fig \ref{fig:PRL:intro}b)).

Using the quasi-static approximation as in the equilibrium case, the steady-state concentration profiles of this two-drop system with radii $R_1$ and $R_2$, such that $R_1 \approx R_2$, are \cite{SI}:
%
%
%
\beqn
\label{eq:PRL:profileIn}
\Pin^{(i)}(r_i) &=& \frac{I_ih}{k+h}+H^{(i)}_{\rm in} \frac{R_i}{r_i} \frac{\sinh(r_i/\xi)}{\sinh (R_i/\xi)}
\\
\label{eq:PRL:profileOut}
\Po^{(i)}(r_i) &=& \frac{Oh}{k+h}+H^{(i)}_{\rm out} \frac{R_i}{r_i} \left( A_i \ee^{r_i/\xi}+B_i \ee^{-r_i/\xi}\right) 
. \ 
\eeqn
In the above, 
 $A_i,B_i$ are independent of $r_i$ {and are given in \cite{SI}}, $I_i$, $O$ denote the combined concentration $P+S$ inside and outside the $i$-th drop, respectively, and are also independent of $r_i$.  Furthermore,
  $H^{(i)}_{\rm in}\equiv \hatPin-I_ih/(k+h)$, $H^{(i)}_{\rm out}\equiv P^{(i)}_{\rm out}(R_i)-Oh/(k+h)$, 
 and $\xi \equiv \sqrt{D/(k+h)}$ is the concentration gradient length scale.
The $S$ profiles are given by $S^{(i)}(r_i)=I_i-\Pin^{(i)}(r_i)$ for $r_i<R_i$ and $S^{(i)}(r_i)=O-\Po^{(i)}(r_i)$ for $r_i>R_i$. 
Note that generally, $O$ is  independent of $r$ {only} when $R_1=R_2$, which we have assumed to be true here as we will focus on the case $R_1 \approx R_2$.

\begin{figure}
	\centering
	\includegraphics[scale=1.02]{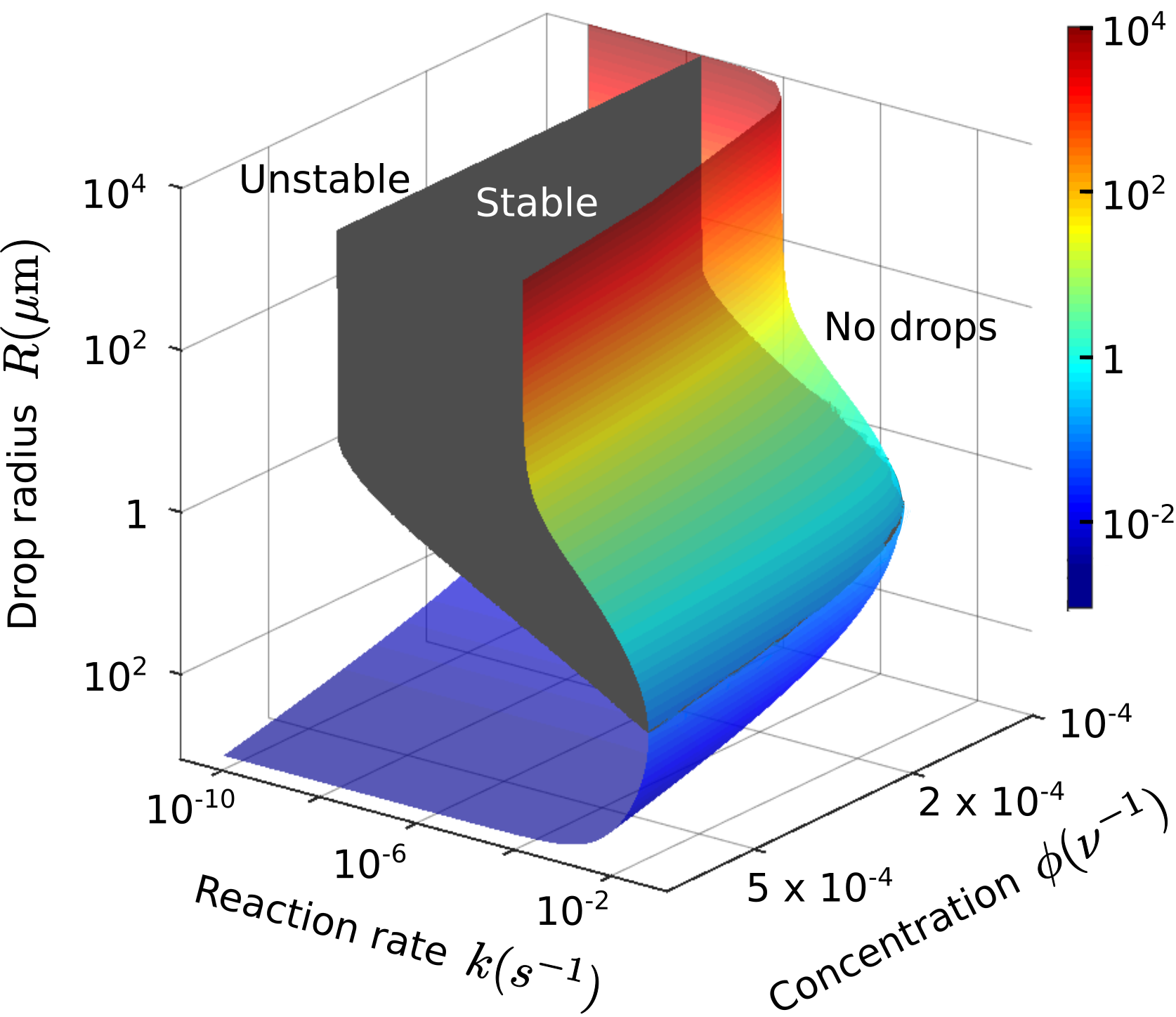}
	\caption{ {\it. The stability of a multi-drop system at fixed backward reaction rate $h$.} The region of existence of a steady-state radius $R^*$ (solution of $g_0(R^*)=0$, \eq \eqref{eq:PRL:g0}) is controlled by the forward reaction rate $k$ and the total solute concentration $\phi$. In the region enclosed by the coloured outer surface, $R^*$ exists and  depends on the drop number density $\rho$ which is not fixed in this figure. The steady-state is stable ($g_1(R^*)<0$, \eq \eqref{eq:PRL:growth:expanded}) above the black inner surface, and unstable ($g_1(R^*)>0$) bellow this surface. Parameters: $h=10^{-2} s^{-1},~ l_c=10^{-2}{\rm \mu m},~ D=1 {\rm \mu m^2 s^{-1}},~\hatPin=5 \times 10^{-2}{ \nu^{-1}},~\hatPo=10^{-4}{\nu^{-1}}$, where $\nu$ is the molecular volume of $P$ and $S$ and can be chosen arbitrarily.
	}
	\label{fig:PRL:surface}
\end{figure}

The volumetric growth rate of the $i$-th  drop in this two-drop system is \cite{zwicker_pnas14} 
\beqn
\label{eq:PRL:growth}
G_i(R_i,R_j) = \frac{4\pi DR_i^2}{\hatPin-\Po^{ (i)}  (R_i)}\left( \left. \frac{\dd \Po^{(i)}}{\dd r_i}\right|_{R_i^{+}}-\left.\frac{\dd \Pin^{(i)}}{\dd r_i} \right|_{R_i^-} \right) .
\ 
\eeqn
Given the drop growth rate above we can study the steady-state drop radius $R^*$ at which the two drops of the same size are in the steady-state ($G_i=0$).
%

We can also analyse its stability by calculating the drops' growth rates upon perturbing their sizes: $R_1\mapsto R^*+\epsilon$ and $R_2\mapsto R^*-\epsilon$. Performing a linear stability analysis, we take $\epsilon\ll R^*$ and expand the growth rate with respect to $\epsilon$:
\beq
\label{eq:PRL:growth:expanded}
G_1(R_1,R_2)= g_0(R^*)+g_1 (R^*) \epsilon +\cO(\epsilon )
\ .
\eeq
Solving for $g_0(R^*)=0$ gives 
the steady-state drop radius $R^*$ and the sign of $g_1 (R^*)$ indicates the stability of the system: coarsening will occur if $g_1>0$ while the system is stable if $g_1<0$. Using the profiles \eqref{eq:PRL:profileIn} \& \eqref{eq:PRL:profileOut},  we find
\beqn
\label{eq:PRL:g0}
&&g_0(R^*) =\frac{4 \pi D R^*}{\hatPin-\Po { (R^*)}} \left[ \Hin \left(1-\frac{R^*}{\xi} \coth{\frac{R^*}{\xi}}\right) \right.\\
\nonumber
&&
\ \ \ \ \ \
-\left .\Ho \left(A\left(1-\frac{R^*}{\xi}\right)e^{\frac{R^*}{\xi}}+B\left(1+\frac{R^*}{\xi}\right)e^{-\frac{R^*}{\xi}}\right) \right] 
\eeqn
with $H_{\rm in/out}\equiv H^{(i)}_{\rm in/out}(R_1=R_2=R^*)$ and $A$ and $B$ are function of $R^*/\xi$ and $L/\xi$ \cite{SI}. 
The expression of $g_1(R^*)$ is more complicated and is shown in 
\cite{SI}.

The surface plot in \fig \ref{fig:PRL:surface} shows for a fixed backward reaction rate $h$, the region of existence of the steady-state radius $R^*$, delimited by the coloured outer surface. Above the black inner surface, the system consists of stable monodisperse drops whose sizes are controlled by the rate $k$, the solute concentration $\phi$ and the drop number density $\rho$ (not fixed in \fig \ref{fig:PRL:surface}). Outside the stable region but still within the outer surface, the monodisperse system is in an unstable steady-state and drops coarsen via Ostwald ripening. Outside the outer surface, drops always shrink.

Interestingly, there are qualitative changes in the system's behaviour as $k$ varies with fixed $h$ as shown in \fig \ref{fig:PRL:compare}, which describes multi-drop stability at fixed solute concentration $\phi$. When $k<k_l$ (blue arrow), the system is in the Lifshitz-Slyozov regime and coarsen (upward arrows), while for $k_l<k<k_u$ (green arrow), the system can be stable (grey region), with co-existing drops of radius determined by $\rho$. In other words, {\it $k_l$ is the critical rate beyond which Ostwald  ripening is arrested.}
Between $k_u$ and $k_c$ (black arrow), the system can also be stable, but with 
an upper bound on the  radius. Beyond $k_c$, no drops can exist in the system  as all drops evaporate  (downward arrows). 

\begin{figure}[]
	\centering
	\includegraphics[scale=1]{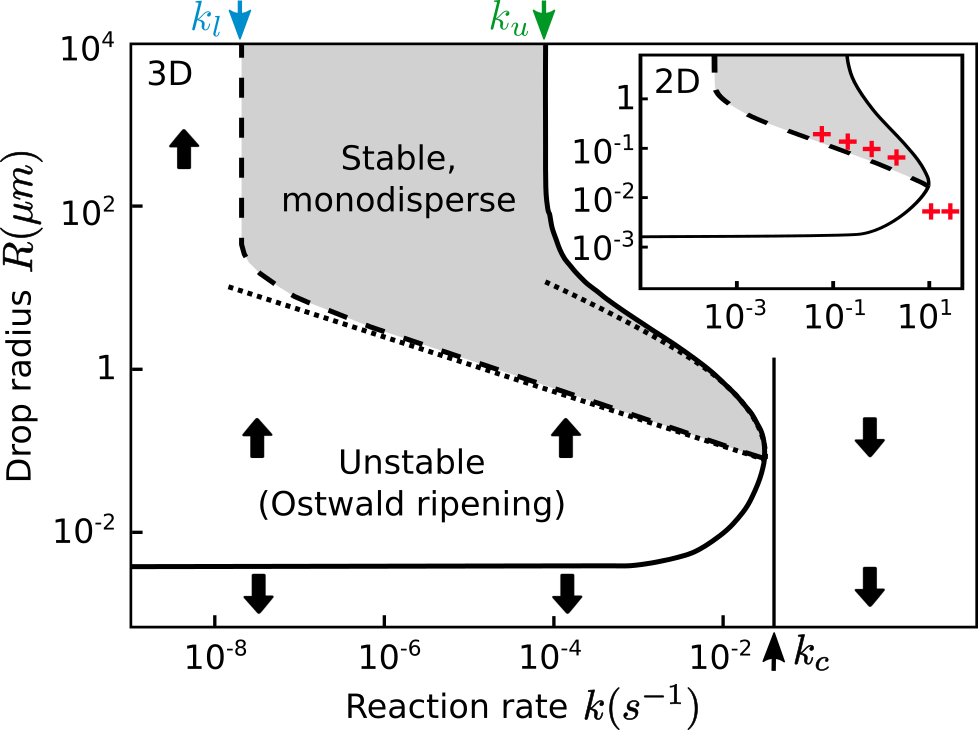}
	\caption{ 
		{\it Stability diagram of a multi-drop system at fixed backward rate $h$ and fixed solute concentration $\phi$.} A steady-state drop radius $R^*$ 
		exists in the region enclosed by the continuous line and depends on the rate $k$ and the drop number density $\rho$. Outside this region no steady-states exist and drops dissolve (downward arrows). The lower part of this line represents the smallest possible drop, or nucleus. 
		Outside the grey region but still within the continuous line the steady-state is unstable 
		causing the average radius to increase (upward arrows). The stability-instability boundary ($g_1(R^*)=0$) is shown with a dashed line.
		 There is a good agreement between our analytical calculation and the numerical solutions  for $k_l,k_u, k_c$.	
		The analytical expressions for the upper bound radius $R_u$ (\eq \eqref{eq:PRL:Ru}) and the stability-instability boundary $R_l$ (\eq \eqref{eq:PRL:Rl}) in the small drop regime ($R^*\ll\xi$) are shown by the dotted lines.
		 Parameters: $\phi=5\times 10^{-4} \nu^{-1}, ~\hatPin=10^{-1}\nu^{-1}$ and the rest are as in \fig \ref{fig:PRL:surface}. 
		{\it Insert:} Comparison between 2D Monte Carlo  simulations and numerical  solutions to the linear stability analysis. Simulation data are shown in red, note that the two rightmost crosses represent the size of the lattice site ($\sim10^{-2} \rm{\mu m}$), i.e., there are no drops in the system. The region $k\approx k_l$ is also investigated in \cite{SI}. See \cite{SI} for the corresponding analysis in 2D and simulation details.\\
		\label{fig:PRL:compare}
	}
\end{figure}

So far, our calculation has been based on our two-drop system with the mean-field matching assumption at the system boundaries. To test this
assumption, we perform Monte Carlo simulations of our ternary model on a 2D lattice with multiple drops to detect the stability-instability boundary 
(black inner surface in \fig \ref{fig:PRL:surface} and dashed curve in \fig \ref{fig:PRL:compare}) and compare the results with our predictions (see \cite{SI} for simulation details.)
The good agreements  are shown in the inset of \fig \ref{fig:PRL:compare} and in \cite{SI}.
%
%

We will now explain analytically the salient features of the stability diagram  by focusing on distinct limits  in the small supersaturation limit.

\textit{Upper bound on drop radius.} We have seen that if $k_u<k<k_c$, there is an upper bound on the drop radius.  We focus here on the regime  $R^*\ll\xi$, which we will see is indeed the case when $k \gg k_u$.
We first analyse the limit of small drop number density $\rho$ so that the distance between drops is large: $L \gg \xi$. By expanding $g_0$ with respect to the small parameters $R^*/\xi$ and $\xi/L$ we seek the set of $R^*$ such that the solutions to $g_0(R^*)=0$ 
 cease to exist. We find
that the expression of this boundary is \cite{SI}:
\beqn
\label{eq:PRL:Ru}
R_u  &=& \sqrt{ \frac{3D \left(\frac{h\phi}{k+h}-\hatPo\left( 1+\frac{l_c}{R_u} \right) \right)}{k \hatPin} } \\
\label{eq:PRL:Ru_bis}
  &\underrel{R_u \gg l_c}{\simeq}& \sqrt{ \frac{3D \left(\frac{h\phi}{k+h}-\hatPo\right)}{k \hatPin} } \ .
\eeqn
which is indicated by the upper dotted line in \fig \ref{fig:PRL:compare}. We have thus recovered the result \eq \eqref{eq:PRL:Ru:intuitive}  obtained by intuitive arguments since $P_{\rm tot}=h\phi/(k+h)$ \cite{SI}.

{\it Stability-instability boundary.}
For the stability-instability boundary we consider large $\rho$ so that $L\ll\xi$. In this case, the small parameters are $R^*/\xi$ and $L/\xi$.
By expanding $g_0$ and $g_1$ around these small parameters, we solve for the steady-state $g_0(R^*)=0$ and then seek the boundary of stability by looking at $g_1(R^*)=0$. The functional form of this boundary is \cite{SI}:
\beq
\label{eq:PRL:Rl}
R_l  = \left( \frac{3 D \l_c \hatPo}{2 k \hatPin}\right)^{\frac{1}{3}}
\ .
\eeq
which is indicated by the lower dotted line in \fig \ref{fig:PRL:compare}. We note that similar scaling laws to \eqs
(\ref{eq:PRL:Ru_bis}) \& 
(\ref{eq:PRL:Rl})  have previously been found for binary mixtures  \cite{zwicker_pre15}.

{\it Critical reaction rate $k_c$.}
The rate $k_c$ beyond which drops dissolve is the solution of $R_u(k_c)=0$ and is maximally bounded as follow \cite{SI}:
\beq
\label{eq:PRL:kc}
k_c < {\rm min} \left[\frac{\phi-\hatPo}{\hatPo} h\ ;~ \frac{4 D \left(\phi-\hatPo\right)^3}{9 l_c^2\hatPin\hatPo^2} \right]
\ . 
\eeq
Note that $k>(\phi-\hatPo)/\hatPo h$ corresponds to the situation where the conversion $P\to S$ is so strong that the system is outside the equilibrium  phase-separating region ($P_{\rm tot}<\hatPo$, see \fig \ref{fig:PRL:intro}a)). 


\textit{Lower and upper critical rates ($k_l$ \& $k_u$).}
Here we focus on the large drop limit so that the small parameters are $\xi/R^*$ and $\xi/L$ (since $L>R^*$). By expanding $g_0, g_1$ with respect to these two small parameters, we solve again for $g_0=0$ and investigate the corresponding stability by looking at $g_1$. Specifically, we find \cite{SI}:
\beq
\label{eq:PRL:kl,ku}
k_l = \frac{2 \l_c \hatPo}{D^{\frac{1}{2}} \hatPin}h^{\frac{3}{2}}  \sep 
 k_u = \frac{2(\phi - \hatPo)}{\hatPin} h
  \ ,
\eeq
where $k_l$ is the transition rate from the stable to the unstable regime, and $k_u$ is the rate beyond which drops' radii have an upper bound.
Thus, the transition to the non-equilibrium regime, namely the arrest of Ostwald ripening, occurs at non-zero reaction rate ($k_l$). This behavior has never been reported before in this system.

Finally, we note that given the richness of the system's behaviour, the generic features of the stability diagram can vary according to $h$, which we have explored in \cite{SI} and in the context of cellular response to environmental stresses \cite{wurtz_a17b}.

In summary, we have studied a phase-separating ternary fluid mixture with chemically active drops. 
 We have categorised the qualitative behavior of the system into distinct regimes based on the reaction rates using a combination of analytical, numerical, and simulation methods. 
 Our work is of direct importance to cytoplasmic organisation, and is also relevant to the control of emulsions in the engineering setting. Interesting future directions include 
 the incorporation of drop coalescence into our coarsening picture, the study of potential shape instabilities in chemically active drops \cite{zwicker_natphys16}, and the generalization of our formalism to many-component mixtures \cite{sear_prl03,jacobs_biophysj17}.

\newpage
\widetext

\setcounter{equation}{0}
\setcounter{figure}{0}

\begin{center}
	
	\textbf{\large Supplemental Materials:\\ \vspace{0.5cm} Chemical reaction-controlled phase separated drops:\\
		Formation, size selection, and coarsening}
	
\end{center}

\renewcommand\theequation{S\arabic{equation}}
\renewcommand\thefigure{S\arabic{figure}}

\part{General theory}

\section{Concentration Profiles and drop growth rates}
In the two-drop system the reaction diffusion equations  in the quasi-static approximation are (see \eqs \eqref{eq:PRL:reacDiffP},\eqref{eq:PRL:reacDiffS} in main text):
\beqn
\label{eq:PRL:SI:reacDiffPin}
0&=D \nabla^2 \Pin^{(i)}(r_i)-k\Pin^{(i)}(r_i)+h\Sin^{(i)}(r_i) \quad\quad &0\leq r_i\leq R_i \\
\label{eq:PRL:SI:reacDiffSin}
0&=D \nabla^2 \Sin^{(i)}(r_i)+k\Pin^{(i)}(r_i)-h\Sin^{(i)}(r_i) \quad\quad &0\leq r_i\leq R_i
\eeqn
and
\beqn
\label{eq:PRL:SI:reacDiffPout}
0&=D \nabla^2 \Po^{(i)}(r_i)-k\Po^{(i)}(r_i)+h\So^{(i)}(r_i) \quad\quad &R_i\leq r_i\leq L \\
\label{eq:PRL:SI:reacDiffSout}
0&=D \nabla^2 \So^{(i)}(r_i)+k\Po^{(i)}(r_i)-h\So^{(i)}(r_i) \quad\quad &R_i\leq r_i\leq L
\eeqn
with $i=1,2$ the drop label, $R_i$ the radius of the $i$-th drop and $L$ the radius of each sub-system (see main text and \fig \ref{fig:PRL:intro}b)). In a multi-drop system $2L$ corresponds to the mean separation between drops and is related to the drop number density $\rho$:
\beqn
\rho=\frac{3}{4 \pi L^3}\ . 
\eeqn
We denote the total solute concentration $P_{\rm tot}+S_{\rm tot}$ by $\phi$ where
$P_{\rm tot}$, $S_{\rm tot}$ are the total concentration of $P$ and $S$, respectively. When phase separation does not occur, the system is homogeneous ($\nabla^2P=\nabla^2S=0$),  and by taking the volume integrals of \eqs \eqref{eq:PRL:SI:reacDiffPin}-\eqref{eq:PRL:SI:reacDiffSout} over the whole system we have
\beqn
\label{eq:PRL:SI:Ptot}
P_{\rm tot}&=&\frac{\phi}{1+\chi} \\
\label{eq:PRL:SI:Stot}
S_{\rm tot}&=&\frac{\chi \phi}{1+\chi}
\ ,
\eeqn
with $\chi \equiv k/h$. When phase separation  occurs and the system is at the steady-state, the concentration gradients must match eactly at the interface. Therefore the diffusion terms cancel out in the volume integrals of \eqs \eqref{eq:PRL:SI:reacDiffPin}-\eqref{eq:PRL:SI:reacDiffSout} and we recover \eqs \eqref{eq:PRL:SI:Ptot}-\eqref{eq:PRL:SI:Stot}. Later we will focus our analysis on small deviations from the steady-state and will approximate $P_{\rm tot}, S_{\rm tot}$ by \eqs \eqref{eq:PRL:SI:Ptot}-\eqref{eq:PRL:SI:Stot}.
Adding \eqs \eqref{eq:PRL:SI:reacDiffPin} + \eqref{eq:PRL:SI:reacDiffSin} and \eqs \eqref{eq:PRL:SI:reacDiffPout} + \eqref{eq:PRL:SI:reacDiffSout} gives
\beqn
\nabla^2\left(P_{\rm in/out}^{(i)}(r_i)+S_{\rm in/out}^{(i)}(r_i)\right)=0
\eeqn
which we solve for two or three spatial dimensions, with spherical or circular symmetry, respectively:
\beqn
P_{\rm in/out}^{(i)}(r_i)+S_{\rm in/out}^{(i)}(r_i)=
\left \lbrace
\begin{array}{l l}
	\frac{a_{\rm in/out}^{(i)}}{r_i}+b_{\rm in/out}^{(i)} & d=3\\
	a_{\rm in/out}^{(i)}\ln{r_i}+b_{\rm in/out}^{(i)} & d=2
\end{array}
\right.
\eeqn
with $a_{\rm in/out}^{(i)}$ and $b_{\rm in/out}^{(i)}$ constants, and $d=2,3$ is the number of spacial dimensions. Inside the drops (``in''), 
the total concentration $\Pin(r_i)+\Sin(r_i)$ must not diverge in the drop center ($r_i=0$), therefore $a_{\rm in}^{(i)}=0$ and $\Pin(r_i)+\Sin(r_i)$ is equal to a constant $I_i$:
\beqn
I_i\equiv P_{\rm in}^{(i)}(r_i)+S_{\rm in}^{(i)}(r_i).
\eeqn
Outside the drop (phase ``out'') and if $R_1=R_2$ the total concentration $\Po(r_i)+\So(r_i)$ must be continuous at the boundary between the two sub-systems ($r_i=L$) therefore $a^{(i)}_{\rm out}=0$. In our study we will focus on small differences in drop radii ($R_1 \approx R_2$) and we make the approximation that $a^{(i)}_{\rm in/out}$ remains zero. Therefore $\Po(r_i)+\So(r_i)$ is equal to a constant $O$ in both sub-systems:
\beqn
O\equiv P_{\rm out}^{(i)}(r_i)+S_{\rm out}^{(i)}(r_i) \quad \quad i=1,2.
\eeqn
We can express $I_i$ and $O$ in terms of the concentrations at the drops' interfaces ($r_i=R_i$):
\beqn
\label{eq:PRL:profile:constant}
I_i&=&\Pin^{(i)}(R_i)+S_{\rm in}^{(i)}(R_i) \\
O&=&\Po^{(i)}(R_i)+S_{\rm out}^{(i)}(R_i)
\eeqn
Using this result the reaction-diffusion systems (\eqs (\ref{eq:PRL:SI:reacDiffPin})-(\ref{eq:PRL:SI:reacDiffSin})) and (\eqs (\ref{eq:PRL:SI:reacDiffPout})-(\ref{eq:PRL:SI:reacDiffSout})) decouple:
\beqn
\label{eq:PRL:SI:decoupledIn}
D\nabla^2 P_{\rm in}^{(i)}(r)-(k+h)P_{\rm in}^{(i)}(r)+hI_i&=&0 \\
\label{eq:PRL:SI:decoupledOut}
D\nabla^2 P_{\rm out}^{(i)}(r)-(k+h)P_{\rm out}^{(i)}(r)+hO&=&0 
\eeqn
and $\Sin^{(i)}(r)=I_i-\Pin^{(i)}(r)$, $\So^{(i)}(r)=O-\Po^{(i)}(r)$. The concentrations and their gradients must be continuous at the sub-system boundaries ($r_i=L$) and we assume the Gibbs-Thomson relations hold at the interface ($r_i=R_i$). This gives the following boundary conditions
\beqn
\nabla_{\br_1} \Po^{(1)}|_{r_1=L} &=& - \nabla_{\br_2} \Po^{(2)} |_{r_2=L} \\
\Po^{(1)}(L) &=& \Po^{(2)}(L) \\
\label{eq:PRL:SI:GibbsThomsonIn}
\Pin^{(i)} (R_i) &=& \hat{P}_{\rm in} \\
\label{eq:PRL:SI:GibbsThomsonOut}
\Po^{(i)} (R_i) &=& \hat{P}_{\rm out} \left(1+l_c/R_i\right).
\eeqn
where $\hatPin$ and $\hatPo$ are the equilibrium coexistence concentrations of $P$ at the interface (see main text \fig \ref{fig:PRL:intro}a)) and $l_c$ is the capillary length. We solve the system \eqs \eqref{eq:PRL:SI:decoupledIn}-\eqref{eq:PRL:SI:GibbsThomsonOut} here in spherical symmetry ($d=3$) or circular symmetry ($d=2$):
\beqn
\label{eq:PRL:SI:profilePin}
\Pin^{(i)}(r)&=&\frac{I_i}{1+\chi} + \Hin^{(i)} \frac{\fin(r)}{\fin(R_i)} \\
\label{eq:PRL:SI:profilePout}
\Po^{(i)}(r)&=&\frac{O}{1+\chi}+ \Ho^{(i)}\left( A_i \fo^{(A)}(r) +  B_i \fo^{(B)}(r)\right) \\
\label{eq:PRL:SI:profileSin}
\Sin^{(i)}(r)&=&I_i-\Pin^{(i)}(r) \\
\label{eq:PRL:SI:profileSout}
\So^{(i)}(r)&=&O-\Po^{(i)}(r)
\ ,
\eeqn
with 
\beqn
\label{eq:PRL:SI:HinDef}
\Hin^{(i)}&\equiv& \Pin^{(i)}(R_i) - \frac{ I_i}{1+\chi} \\
\label{eq:PRL:SI:HoutDef}
\Ho^{(i)}&\equiv& \Po^{(i)}(R_i) - \frac{O}{1+\chi}
\ ,
\eeqn
and
\beqn
\label{eq:PRL:SI:fInOutd2}
\fin(r)\equiv {\rm Re}\left[J_0\left(\iota\frac{r}{\xi}\right)\right], \quad
\fo^{(A)}(r)\equiv {\rm Re}\left[J_0\left(\iota\frac{r}{\xi}\right)\right], \quad
\fo^{(B)}(r)\equiv {\rm Re}\left[Y_0\left(-\iota\frac{r}{\xi}\right)\right] \ ,\quad\quad d=2 \\ \nonumber  \\
\label{eq:PRL:SI:fInOutd3}
\fin(r)\equiv\frac{R}{r}\sinh{\frac{r}{\xi}} , \quad
\fo^{(A)}(r)\equiv\frac{R}{r} e^{r/\xi}, \quad
\fo^{(B)}(r)\equiv\frac{R}{r} e^{-r/\xi} \ , \quad\quad d=3 \ .
\eeqn
$\xi\equiv\sqrt{D/(k+h)}$ is the gradient length scale, $J_0$ and $Y_0$ are the 0-th order Bessel functions of the first and second kind, respectively, and $\iota$ is the imaginary unit $\sqrt{-1}$.
$A_i$ and $B_i$ are independent of $r_i$ and are solutions of the system:
\beqn
\label{eq:PRL:SI:BC:AB1}
A_i \fo^{(A)}(R_i)+B_i \fo^{(B)}(R_i)&=&1 \quad\quad\quad i=1,2\\
\label{eq:PRL:SI:BC:AB2}
\Ho^{(1)}\left(A_1\fo^{(A)}(L)+B_1\fo^{(B)}(L)\right)&=&\Ho^{(2)}\left(A_2\fo^{(A)}(L)+B_2 \fo^{B)}(L) \right) \\
\label{eq:PRL:SI:BC:AB3}
\Ho^{(1)}\left( A_1 \fo^{(A)'}(L)+B_1\fo^{(B)'}(L)\right)&=&-\Ho^{(2)}\left( A_2 \fo^{(A)'}(L)+B_2\fo^{(B)'}(L)\right).
\eeqn
The $i$-th drop volumetric growth is \cite{zwicker_pnas14_bis}:
\beqn
G^{(i)}(R_i,R_j)& \equiv& \frac{4\pi DR_i^2}{\hatPin-\Po^{(i)}(R_i)}\left( \left. \frac{\dd \Po^{(i)}}{\dd r_i}\right|_{R_i^+}-\left.\frac{\dd \Pin^{(i)}}{\dd r_i} \right|_{R_i^-} \right) \sep j\neq i \nonumber\\
\label{eq:PRL:SI:growth}
&=&\frac{4\pi DR_i^2}{\hatPin-\Po^{(i)}(R_i)}\left[\Ho^{(i)}\left(A_i \fo^{(A)'}(R_i)+B_i \fo^{(B)'}(R_i)\right) - \Hin^{(i)} \frac{\fin^{'}(R_i)}{\fin(R_i)} \right] \ .
\eeqn
\section{Concentration jump of S at the drop interface}
We denote the discontinuity of the concentration $S$ at the interface by $\Delta S$:
\beqn
\Delta S \equiv \frac{\Sin^{(i)}(R_i)}{\So^{(i)}(R_i)} \ .
\eeqn
We impose the conservation of the total number of molecules in the system:
\beqn
I_1 R_1^d+I_2 R_2^d+O\left(2 L^d-R_1^d-R_2^d\right)=2 \phi L^d,
\eeqn
leading to
\beqn
\label{eq:PRL:SI:Sin}
\Sin^{(i)}(R_i)&=&\Delta S \frac{\phi-\Po^{(i)}(R_i)-\frac{1}{2}\left[ \left(\hatPin-\Po^{(1)}(R_1)\right)\left(\frac{R_1}{L}\right)^d + \left(\hatPin-\Po^{(2)}(R_2)\right)\left(\frac{R_2}{L}\right)^d \right]}{1-\frac{1}{2}\left(\frac{R_j}{L}\right)^d + \frac{1}{2}\Delta S \frac{R_1^d+R_2^d}{L^d}} \ , \  j\neq i\\ 
\label{eq:PRL:SI:Sout}
\So^{(i)}(R_i)&=&\frac{\Sin^{(i)}(R_i)}{\Delta S}
\eeqn
The profiles (\eqs \eqref{eq:PRL:SI:profilePin}-\eqref{eq:PRL:SI:profileSout}) are now fully defined as functions of $R_1, R_2, \Delta S$. 

\section{Steady-state drop radius $R^*$}
A system with identical drop radii $R^*$ is at steady-state if the drop growths $G^{(i)}(R^*,R^*)$ (\eq \eqref{eq:PRL:SI:growth}) are zero. Therefore the steady-state condition is
\beqn
\label{eq:PRL:SI:steady:general}
\Ho\left(A \fo^{(A)'}(R^{*}) + B \fo^{(B)'}(R^{*})\right)-\Hin \frac{\fin^{'}(R^{*})}{\fin(R^{*})}=0
\ ,
\eeqn
where $\fo^{(A)'}$ denotes the derivative of $\fo^{(A)}$, etc, and
$A\equiv A_i(R^*,R^*)$, $B\equiv B_i(R^*,R^*)$, $\Hin\equiv \Hin^{(i)}(R^*,R^*)$ and $\Ho\equiv \Ho^{(i)}(R^*,R^*)$. Using $R_1=R_2=R^*$, the system \eqs (\ref{eq:PRL:SI:BC:AB1})-(\ref{eq:PRL:SI:BC:AB3}) reduces to
\beqn
\label{eq:PRL_SI_BC_steady}
A \fo^{(A)}(R^{*})+ B \fo^{(B)}(R^{*})&=&1 \\
A \fo^{(A)'}(L)+ B \fo^{(B)'}(L)&=&0
\eeqn
and we solve for $A$ and $B$:
\beqn
\label{eq:PRL:SI:A:general}
&&A=\frac{\fo^{(B)'}(L)}{\fo^{(A)}(R^{*}) \fo^{(B)'}(L)-\fo^{(A)'}(L) \fo^{(B)}(R^{*})} \\
\label{eq:PRL:SI:B:general}
&&B=-\frac{\fo^{(A)'}(L)}{\fo^{(A)}(R^{*}) \fo^{(B)'}(L)-\fo^{(A)'}(L) \fo^{(B)}(R^{*})}
\eeqn
Plugging \eqs (\ref{eq:PRL:SI:Sin}) and (\ref{eq:PRL:SI:Sout}) for $R_1=R_2=R^*$ in the definitions of $H_{\rm in/out}$ (\eqs \eqref{eq:PRL:SI:HinDef}, \eqref{eq:PRL:SI:HoutDef})  we find
\beqn
\label{eq:PRL:SI:Hin}
\Hin&\equiv \Hin^{(i)}(R^*,R^*)= &\Pin-\frac{ \Delta S \phi + \left(\hatPin-\Delta S\Po(R^{*}) \right)\left(1-\left(\frac{R}{L}\right)^d\right)}{(\chi+1)\left(1-(1-\Delta S)\left(\frac{R}{L}\right)^d\right)} \\
\label{eq:PRL:SI:Hout}
\Ho&\equiv \Ho^{(i)}(R^*,R^*)= &\Po-\frac{\phi-\left(\hatPin-\Delta S \Po(R^{*}) \right)\left(\frac{R}{L}\right)^d}{(\chi+1)\left(1-(1-\Delta S)\left(\frac{R}{L}\right)^d\right)}.
\eeqn
where we have dropped the unnecessary upper script $(i)$ in $\Po$,

\section{Linear stability of the steady-state}
\label{sec:PRL:SI:stability}
We perturb the drop sizes about the steady-state:
\beqn
R_1&=&R^{*} +\epsilon \\
R_2&=&R^{*} -\epsilon \\
\eeqn
with $\epsilon \ll R^{*}$. We focus on the growth rate of the drop 1 ($G^{(1)}$). Expanding for the small parameter $\epsilon/R^{*}$:
\beqn
G^{(1)}(R_1,R_2)=g_0(R^{*}) + \epsilon g_1(R^{*}) + \cO\left(\epsilon^2\right)
\eeqn
with
\beqn
g_0(R^*)&\equiv & G^{(1)}(R^*,R^*)=G^{(2)}(R^*,R^*)\\
g_1(R^{*})& \equiv &\left. \frac{\partial G^{(1)}}{\partial R_1}\right|_{R^{*}, R^{*}} - \left. \frac{\partial G^{(1)}}{\partial R_2}\right|_{R^{*}, R^{*}} \ ,
\eeqn
we find
\beqn
g_0(R^*)&=&0 \quad \Longleftrightarrow \quad \Ho\left(A \fo^{(A)'}(R^{*}) + B \fo^{(B)'}(R^{*})\right)-\Hin \frac{\fin^{'}(R^{*})}{\fin(R^{*})}=0 \\
\label{eq:PRL:SI:g1:general}
\frac{g_1(R^{*})}{4 \pi D R^{*2}}&=&\HHo \left[ A \foAp(R^{*}) + B \foBp(R^{*}) \right] + \Ho \left[ \mathcal{A} \foAp(R^{*}) + \mathcal{B} \foBp(R^{*})+A \foApp(R^{*}) + B \foBpp(R^{*})\right] \nonumber \\
&&-\HHin \frac{\fin^{'}(R^{*})}{\fin(R^{*})}- \Hin \left[ \frac{\fin^{''}(R^{*})}{\fin(R^{*})}-\left( \frac{\fin^{'}(R^{*})}{\fin(R^{*})}\right)^2\right] \ ,
\eeqn
with $\mathcal{A}\equiv \left.\frac{\partial A_1}{\partial R_1}\right|_{R^{*}, R^{*}}-\left.\frac{\partial A_1}{\partial R_2}\right|_{R^{*}, R^{*}}$, $\mathcal{B}\equiv \left.\frac{\partial B_1}{\partial R_1}\right|_{R^{*}, R^{*}}-\left.\frac{\partial B_1}{\partial R_2}\right|_{R^{*}, R^{*}}$ and $\mathcal{H}_{\rm in/out}\equiv \left.\frac{\partial H^{(1)}_{\rm in/out}}{\partial R_1}\right|_{R^{*}, R^{*}}-\left.\frac{\partial H^{(1)}_{\rm in/out}}{\partial R_2}\right|_{R^{*}, R^{*}}$.
The steady-state radius $R^*$ is obtained by solving $g_0(R^*)=0$ and the sign of $g_1(R^*)$ indicates the stability of the steady-state (stable if $g^{(1)}<0$ and unstable if $g^{(1)}>0$). 
Expanding $A_i$ for small $\epsilon/R^*$:
\beqn
A_1(R_1,R_2)&=&A+ \epsilon \left( \left.\frac{\partial A_1}{\partial R_1}\right|_{R^{*}, R^{*}}-\left.\frac{\partial A_1}{\partial R_2}\right|_{R^{*}, R^{*}} \right) + \cO\left(\epsilon^2\right)\\
A_2(R_1,R_2)&=&A + \epsilon \left( \left.\frac{\partial A_2}{\partial R_1}\right|_{R^{*}, R^{*}}-\left.\frac{\partial A_2}{\partial R_2}\right|_{R^{*}, R^{*}} \right) + \cO\left(\epsilon^2\right)
\eeqn
The two-drop system must be unchanged by the permutation of the two drops, therefore
\beqn
\left.\frac{\partial A_2}{\partial R_1}\right|_{R^{*}, R^{*}}=\left.\frac{\partial A_1}{\partial R_2}\right|_{R^{*}, R^{*}} \\
\left.\frac{\partial A_1}{\partial R_1}\right|_{R^{*}, R^{*}}=\left.\frac{\partial A_2}{\partial R_2}\right|_{R^{*}, R^{*}},
\eeqn
and it follows that
\beqn
A_1(R_1,R_2)&=&A + \epsilon \mathcal{A} + \cO\left(\epsilon^2\right) \\
A_2(R_1,R_2)&=&A - \epsilon \mathcal{A} + \cO\left(\epsilon^2\right).
\eeqn
This can be generalized for the other quantities:
\beqn
B_1(R_1,R_2)&=&B + \epsilon \mathcal{B} + \cO\left(\epsilon^2\right) \\
B_2(R_1,R_2)&=&B - \epsilon \mathcal{B} + \cO\left(\epsilon^2\right) \\
\Hin^{(1)}(R_1,R_2)&=&\Hin + \epsilon \HHin + \cO\left(\epsilon^2\right) \\
\Hin^{(2)}(R_1,R_2)&=&\Hin - \epsilon \HHin + \cO\left(\epsilon^2\right) \\
\Ho^{(1)}(R_1,R_2)&=&\Ho + \epsilon \HHo + \cO\left(\epsilon^2\right) \\
\Ho^{(2)}(R_1,R_2)&=&\Ho - \epsilon \HHo + \cO\left(\epsilon^2\right) \\
\eeqn
Using these results the boundary condition system \eqs (\ref{eq:PRL:SI:BC:AB1})-(\ref{eq:PRL:SI:BC:AB3}) reduces to
\beqn
A \foAp(R^{*})+\mathcal{A}\foA(R^{*})+B \foBp(R^{*}) +\mathcal{B}\foB(R^{*})&=&0\\
\Ho\left[\mathcal{A}\foA(L)+\mathcal{B}\foB(L)\right]+\HHo \left[ A \foA(L)+B\foB(L) \right]&=&0
\eeqn
and we can solve for $\mathcal{A},\mathcal{B}$:
\beqn
\label{eq:PRL:SI:AA}
\mathcal{A}&=& \frac{\frac{\HHo}{\Ho}\left(A \foA(L)+B\foB(L)\right)\foB(R^{*})-\left(A\foAp(R^{*})+B\foBp(R^{*})\right)\foB(L)}{\foA(R^{*})\foB(L)-\foA(L)\foB(R^{*})} \\
\label{eq:PRL:SI:BB}
\mathcal{B}&=&-\frac{\frac{\HHo}{\Ho}\left(A \foA(L)+B\foB(L)\right)\foA(R^{*})-\left(A\foAp(R^{*})+B\foBp(R^{*})\right)\foA(L)}{\foA(R^{*})\foB(L)-\foA(L)\foB(R^{*})}. 
\eeqn
Plugging \eqs \eqref{eq:PRL:SI:Hin} and \eqref{eq:PRL:SI:Hout} in the definitions of $\mathcal{H_{\rm in/out}}$, we find
\beqn
\HHin&=&-\frac{\Delta S \hatPo l_c}{(\chi+1)R^2} \\
\HHo&=&-\frac{\hatPo l_c}{R^2} 
\ .
\eeqn

\part{Three dimensions: continuous concentration of $S$ at the drop interface ($  \triangle S=1$)}
\setcounter{section}{0}
We will now study analytically the system in the small supersaturation regime, which amounts to $R\ll L$.

\section{Steady-state}
For $d=3$, using \eqs \eqref{eq:PRL:SI:fInOutd3}, the steady-state condition (\eq \eqref{eq:PRL:SI:steady:general}) becomes
\beqn
\label{eq:PRL:SI:steady:d=3}
\Hin \left(y \coth(y)-1\right) + \Ho \left( A(1-y)e^{y}+B(1+y)e^{-y} \right) =0
\eeqn
with $x\equiv L/\xi$, $y \equiv R^*/\xi$ and the coefficients $A$ and $B$ (\eqs \eqref{eq:PRL:SI:A:general} and \eqref{eq:PRL:SI:B:general}) are
\beqn
\label{eq:PRL:SI:A:d=3}
&&A=\frac{(x+1)e^{-x}}{(x-1)e^{x-y}+(x+1)e^{-(x-y)}}  \\
\label{eq:PRL:SI:B:d=3}
&&B=\frac{(x-1)e^{x}}{(x-1)e^{x-y}+(x+1)e^{-(x-y)}} \ .
\eeqn
We decouple $x$ and $y$ in \eq \eqref{eq:PRL:SI:steady:d=3} with the use of the identity $a e^c+be^{-c}=(a+b)\cosh(c)+(a-b)\sinh(c)$:
\beqn
\label{eq:PRL:SI:steadyState}
\frac{\sinh x-x\cosh x}{\cosh x-x\sinh x} = \frac{(\lambda+1)\left(\sinh y -y\cosh y \right)}{\lambda \left(\cosh y -y\sinh y\right)-\cosh y\left(y~{\rm coth}~y-1\right)}
\eeqn
with
\beqn
\lambda &\equiv& -\Ho/\Hin \nonumber \\
\label{eq:PRL:SI:lambda}
&=& \frac{\phi - \Po(R^*) (1+\chi)-\frac{(R^*)^3}{L^3}(\hatPin-\Po(R^*)) }{\hatPin \chi + \Po(R^*) -\phi + \frac{(R^*)^3}{L^3}(\hatPin-\Po(R^*))} 
\ .
\eeqn
\section{Stability}
From \eqs \eqref{eq:PRL:SI:fInOutd3}, \eqref{eq:PRL:SI:g1:general}, \eqref{eq:PRL:SI:AA} and \eqref{eq:PRL:SI:BB} we find
\beqn
g_1&=& 4 \pi D  \left[ \Hin \left(y^2 {\rm csch}^2 y-1\right) +\Ho \left( \left (\mathcal{A}(y-1)e^{y}-\mathcal{B}(1+y)e^{-y}\right)R +2+y^2-2y\left(Ae^y-Be^{-y}\right)\right) \right.\\
&&\left. \quad\quad~ +\mathcal{\Hin}\left(1-y\coth{y}\right)R + \mathcal{\Ho} \left[A(y-1)e^{y}-B(1+y)e^{-y}\right]R \right]
\eeqn
and
\beqn
\mathcal{A}&=&\frac{-\frac{\HHo}{\Ho}\left(Ae^x+Be^{-x}\right)e^{-y}+\frac{1}{R}\left(A(y-1)e^{y}-B(1+y)e^{-y}\right)e^{-x} }{e^{x-y}-e^{-(x-y)}}  \\
\mathcal{B}&=&\frac{\frac{\HHo}{\Ho}\left(Ae^x+Be^{-x}\right)e^{y}-\frac{1}{R}\left(A(y-1)e^{y}-B(1+y)e^{-y}\right)e^{x} }{e^{x-y}-e^{-(x-y)}}.
\eeqn
We rearrange in the more convenient form:
\beq
\label{eq:PRL:SI:g1}
g_1= 4 \pi D \left( f_1 \Hin +  f_2\Ho +  f_3 R \mathcal{\Hin} +  f_4 R \mathcal{\Ho} \right) \ ,
\eeq
where
\beqn
\begin{array}{ll}
	\label{eq:PRL:SI:f1}
	&f_1\equiv y^2 {\rm csch}^2 y-1\\
	\label{eq:PRL:SI:f2}
	&f_2 \equiv  \dfrac{(1-x)e^{2(x-y)}+(1+x)e^{-2(x-y)}+4y(x-y)-2}{(1-x)e^{2(x-y)}+(1+x)e^{-2(x-y)}-2}\\
	\label{eq:PRL:SI:f3}
	&f_3 \equiv 1-y\coth y \\
	\label{eq:PRL:SI:f4}
	&f_4 \equiv -\dfrac{(1+y)e^{x-y}+(y-1)e^{-(x-y)}}{e^{x-y}-e^{-(x-y)}}
	\ .
\end{array}
\eeqn

\begin{figure}
	\centering
	\includegraphics[scale=1]{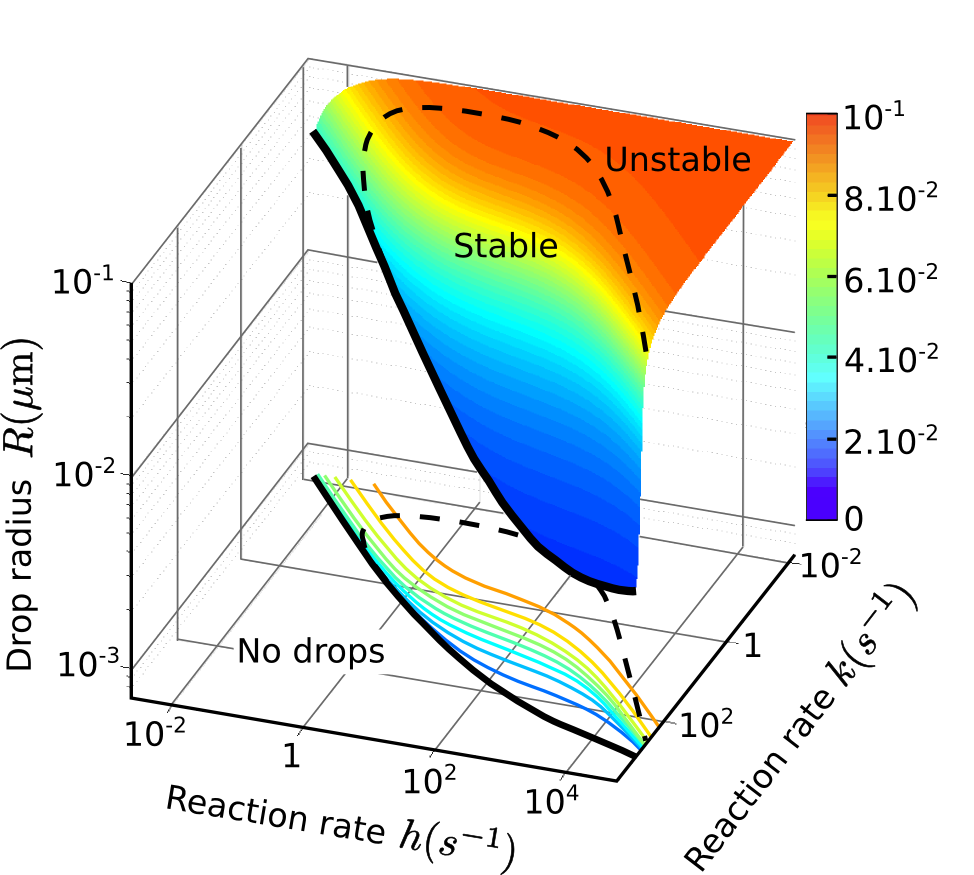}
	\caption{{\it. The stability of a multi-drop system at fixed drop number density $\rho$.} The steady-state radius $R^*$ (solution of $g_0(R^*)=0$, \eq \eqref{eq:PRL:SI:steadyState})  is controlled by the reaction rates $k$ and $h$. The continuous line delimits the region where $R^*$ exists. The steady-state is stable ($g_1(R^*)<0$, \eq \eqref{eq:PRL:SI:g1}) inside the region enclosed by the dashed line and the continuous line, and unstable ($g_1(R^*)>0$) outside this region.  Parameters: $\rho=1 {\rm  \mu m^{-3}},~ \l_c=10^{-2}{\rm \mu m},~ D=1 {\rm \mu m^2 s^{-1}},~\phi=5.10^{-4}{ /\nu},~\hatPin=10^{-1}{ /\nu},~\hatPo=10^{-4}{/\nu}$, where $\nu$ is the molecular volume of $P$ and $S$  and can be chosen arbitrarily.
		\label{fig:PRL:SI:surface}
	}
\end{figure}

The surface plot in \fig \ref{fig:PRL:SI:surface} shows the steady-state radius $R^*$ for a fixed drop number density $\rho$ and the stable region is enclosed by a dashed line.

\section{Equilibrium systems}

When $k=0$ or/and $h=0$ no reactions occur in the steady-state and the system is in equilibrium conditions. From \eqs \eqref{eq:PRL:SI:Ptot}-\eqref{eq:PRL:SI:Stot}, $k=0$ with $h>0$ leads to $P_{\rm tot}=\phi$, $S_{\rm tot}=0$, and $k>0$ with $h=0$ to $P_{\rm tot}=0$, $S_{\rm tot}=\phi$. When $k=h=0$ the ratio $\chi\equiv k/h$ and \eqs \eqref{eq:PRL:SI:Ptot}-\eqref{eq:PRL:SI:Stot} are undefined. We can nonetheless study such systems at concentrations $P_{\rm tot}$, $S_{\rm tot}$ using our non-equilibrium formalism by making $k$ and $h$ converge to zero while keeping $\chi$ in such a way that we recover the desired $P_{\rm tot}, S_{\rm tot}$ from \eqs \eqref{eq:PRL:SI:Ptot}-\eqref{eq:PRL:SI:Stot}:
\beqn
\label{eq:PRL:SI:chi_eq}
\chi=\frac{\phi-P_{\rm tot}}{P_{\rm tot}} \ .
\eeqn
Using this prescription we now calculate the steady-state drop radius $R^*$ and determine its stability. Taking $k,h$ to zero implies that $\xi \rightarrow \infty$ thus $x,y \rightarrow 0$ (but $y/x=R/L$) and therefore the steady-state condition \eq \eqref{eq:PRL:SI:steadyState} becomes
\beqn
(R^*)^3 = \frac{\lambda}{\lambda+1} L^3 \ ,
\eeqn
and using \eqs \eqref{eq:PRL:SI:chi_eq} and \eqref{eq:PRL:SI:lambda} we find
\beqn
(R^*)^3 &=& \frac{P_{\rm tot}-\Po}{\hatPin}  L^3 \left( 1+\cO\left(\frac{\hatPo}{\hatPin}\right) \right) \ . 
\eeqn
Since at equilibrium there are no concentration gradients (this can be seen by taking $k=h=0$ in \eqs \eqref{eq:PRL:SI:reacDiffPin}-\eqref{eq:PRL:SI:reacDiffSout}), this result can also be recovered simply by imposing the conservation of the number of molecules $P$ in the system: $R^3 \Pin + (L^3-R^3)\Po = L^3 P_{\rm tot} $. Plugging the Gibbs-Thomson relation (\eq \eqref{eq:PRL:SI:GibbsThomsonOut}) in this result, we find the influence of the surface tension on the drop's radius:
\beqn
\label{eq:PRL:SI:Req}
(R^*)^3 = \left( \frac{P_{\rm tot}-\hatPo}{\hatPin} -\frac{\hatPo l_c}{\hatPin R^*}  \right) \left(1+\cO\left(\frac{\hatPo}{\hatPin}\right) \right) L^3 \ .
\eeqn
The drop radius thus scales as the system size ($\propto L$) with a negative finite size correction ($\propto 1/R$). We also find the radius $R_n$ of the nucleus, which is the smallest drop that can exist: 
\beqn
\label{eq:PRL:SI:Rc}
R_n \simeq \frac{\hatPo l_c}{P_{\rm tot}-\hatPo} \ .
\eeqn
Smaller drops dissolve because the concentration of $P$ outside the drop next to interface is larger than the total concentration $P_{\rm tot}$. We now study the stability of a multi-drop system by taking $k,h =0$ (leading to $x,y\rightarrow 0$ but $y/x=R/L$) in \eqs \eqref{eq:PRL:SI:f1}:
\beqn
f_1&=& 0 \\
f_2&=& 0 \\
f_3&=& 0 \\
f_4&=& -1+\cO\left(\frac{R}{L}\right) \ ,
\eeqn
and it follows that the stability relation (\eq \eqref{eq:PRL:SI:g1}) is:
\beqn
g_1 \simeq   \frac{4 \pi D \hatPo l_c}{R^*}  >0 \ .
\eeqn
Since $g_1>0$ for all radii $R^*$ we recover the equilibrium result that a multi-drop system is always unstable to Ostwald ripening \cite{lifshitz_jpcs61_bis}.

\section{Non-equilibrium systems}

We now focus on non-equilibrium systems ($k,h>0$). We first expose qualitative arguments showing that drops shrink when chemical reactions are present, then we study quantitatively multi-drop systems in different regimes based on the drop radius $R$ and the inter-drop distance $L$ compared to the gradient length scale $\xi$, and the reaction rates $k,h$.

\subsection{Chemical reactions lead do drop shrinkage and larger critical radius}
\label{sec:PRL:SI:noneq:intro}

We first consider an equilibrum ($k=h=0$) single-drop system with total concentrations $P_{\rm tot}$ and $S_{\rm tot}$ and $\phi \equiv P_{\rm tot}+S_{\rm tot} \ll 1$. The drop size is given by \eq \eqref{eq:PRL:SI:Req}. We then switch on the chemical reactions ($k,h>0$) in such a way that $P_{\rm tot}$ and $S_{\rm tot}$ remain unchanged ($k/h$ is given by \eq \eqref{eq:PRL:SI:chi_eq}). Outside the drop, the concentrations of both $P$ and $S$ are small and we neglect the chemical reactions. Inside the drop however, the $P$ concentration is high so we expect that the reaction $P\rightarrow^k S$ dominates and depletes $P$ from the drop, leading to the drop's shrinkage. From this intuitive argument we expect that drops are smaller when chemical reactions are present  compared to the equilibrium case.

We now show that this argument is indeed correct. At the exact time $t=0$ at which chemical reactions are switched on, the concentration profiles are flat inside and outside the drop and we can predict qualitatively how the system reacts after a small time interval $t=dt$. Using the reaction-diffusion equations (\eqs \eqref{eq:PRL:SI:reacDiffPin}-\eqref{eq:PRL:SI:reacDiffPout}) with $\nabla^2P_{\rm in/out}=0$ we find the variation of the concentration $P$ inside and outside the drop:
\beqn
\frac{dP_{\rm in}}{dt} &=& -k \hatPin + h S_{\rm tot}\\
\frac{dP_{\rm out}}{dt} 	&=&  -k \Po + h S_{\rm tot}  \ ,
\eeqn
and since $S_{\rm tot}= P_{\rm tot}k/h$ (\eq \eqref{eq:PRL:SI:Stot}):
\beqn
\frac{dP_{\rm in}}{dt}  &=& -k \left(\hatPin - P_{\rm tot}\right) \\
\frac{dP_{\rm out}}{dt}  &=& k \left(P_{\rm tot}- \Po\right)   \ .
\eeqn
$\hatPin > P_{\rm tot}> \Po$ is a condition for phase separation to occur due to the conservation of the number of molecules $P$ in the system. Moreover we focus only on systems where the drop density $R^3/L^3$ is small, so from  \eq \eqref{eq:PRL:SI:Req} we must have $P_{\rm tot} \ll \hatPin$. As a result, the decrease in concentration inside the drop must be larger than the increase in concentration outside the drop. Because of the fixed interfacial boundary conditions, we expect the gradient  inside the drop next to the interface to be greater than that right outside the drop, i.e.,
\beq
\left. \frac{\pp P_{\rm in}}{\pp r}\right|_R  > \left. \frac{\pp P_{\rm out}}{\pp r}\right|_R
\ .
\eeq
Therefore, the concentration of $P$ is depleted at the interface, and as a result the drop shrinks.

%

Let us now see the effect of chemical reactions on the critical radius, or nucleus radius (see \eq \eqref{eq:PRL:SI:Rc} for the equilibrium case). Consider a nucleus at equilibrium condition ($k=h=0$) with radius given by \eq \eqref{eq:PRL:SI:Rc}. From the Gibbs-Thomson relation (\eq \eqref{eq:PRL:SI:GibbsThomsonOut}) we know that the $P$ concentration right outside the nucleus is identical to $P_{\rm tot}$. When chemical reactions are turned on (keeping $P_{\rm tot}, S_{\rm tot}$ constant) the nucleus must shrink from the argument we have just exposed, and as a result the concentration of $P$ just outside the nucleus will exceed $P_{\rm tot}$ (\eq \eqref{eq:PRL:SI:GibbsThomsonOut}). This breaks the requirement that the total number of molecules $P$ must be conserved, thus leading to the nucleus dissolution. To compensate for this effect the nucleus in non-equilibrium conditions is necessarily larger than the nucleus at equilibrium.

We have shown qualitatively that when chemical reactions are turned on, drops shrink while the size of the smallest possible drop that can exist, the nucleus, is larger. The evaluation of the steady-state radius is more involved and must account for the chemical reactions-induced concentration gradients (\eqs \eqref{eq:PRL:SI:profilePin}-\eqref{eq:PRL:SI:profileSout}) (see \sect \ref{sec:PRL:SI:largeDrops}-\ref{sec:PRL:SI:smallDropsHighRho}). 

\subsection{Large drops ($R \gg \xi$)}
\label{sec:PRL:SI:largeDrops}
We focus here on the regime where drop radii $R$ are large compared to the gradient length scale $\xi$. Since the inter-drop distance $L$ is always larger than $R$ this regime also implies that $L\gg\xi$. 
\subsubsection{Steady-state}

We expand the steady-state condition \eq \eqref{eq:PRL:SI:steadyState} for $x\gg y \gg 1$ and find:
\beqn
y=\frac{1+\lambda}{1-\lambda} \left(1+\cO\left(\frac{1}{x}\right) +\cO\left(e^{-2(x-y)}\right) + \cO\left(e^{-2y}\right) \right).
\eeqn
For $y\gg1$ to be true we must also have
\beqn
\lambda < 1  , \quad \cO(\lambda)=\cO(1) \ ,
\eeqn
and we further expand using $\lambda-1$ as a small parameter:
\beqn
\label{eq:PRL:SI:steadyLarge1}
y = \frac{2}{1-\lambda}\left(1 +\cO\left(\frac{1}{y}\right)+\cO\left(e^{-2(x-y)}\right)  \right) \ .
\eeqn
Using \eqs \eqref{eq:PRL:SI:lambda} and expanding further in the small parameters $l_c/R^*$ and $\xi/R^*$ we get the steady-state radius $R^*$:
\beqn
\label{eq:PRL:SI:steadyLargeR}
(R^*)^3 &=& \left(a + \frac{b}{R^*}\right) L^3 \ ,
\eeqn
with
\beqn
a&=& \left( \frac{\phi-\hatPo}{\hatPin}-\frac{\chi}{2}\left(1+\cO\left(\frac{1}{y}\right)\right)\right) \left( 1+\cO\left(\frac{\hatPo}{\hatPin}\right)\right) \\
b&=&-\frac{\hatPo l_c}{\hatPin} \left( 1+\cO\left(\frac{\hatPo}{\hatPin}\right)\right) +  \frac{\chi\xi}{2} \left( 1 + \cO\left(\frac{1}{y}\right)+\cO\left(e^{-2(x-y)}\right) \right) \ .
\eeqn
In the large drop limit $b/R^*\rightarrow0$ and there is a critical rate $k_u$ above which drops cease to exist ($R^*<0$):
\beqn
\label{eq:PRL:SI:ku}
k_u = \frac{2(\phi-\hatPo)h}{\hatPin}  \left[ 1+\cO\left(\frac{\hatPo}{\hatPin}\right)\right] \ .
\eeqn
We will later show that drops can still exist for $k>k_u$ but only with radii $R^*$ smaller than the gradient length scale $\xi$ ($y<1$). From \eq \eqref{eq:PRL:SI:steadyLargeR} we also see that $R^*$ scales as the system size $(\propto L)$, with a finite size correction ($b/R$). When $k=0$ and $h>0$ ($\chi=0$) no chemical reactions occur and we recover the equilibrium steady-state radius (\eq \eqref{eq:PRL:SI:Req}) because $P_{\rm tot}=\phi$ (\eq \eqref{eq:PRL:SI:chi_eq}). In particular the finite size correction is negative ($b=-\hatPo l_c /\hatPin $) and originates from the Gibbs-Thomson relation (\eq \eqref{eq:PRL:SI:GibbsThomsonOut}). Interestingly when chemical reactions are switched on ($k,h>0$) the correction becomes positive if the rate $k$ is larger than a critical value which we find by solving $b(k)=0$:
\beqn
\label{eq:PRL:SI:inverseGibbs}
k=\frac{2 l_c \hatPo h^{3/2}}{D^{1/2}\hatPin } \left[1+\cO\left(\frac{k}{h}\right) + \cO\left(\frac{\hatPo}{\hatPin}\right)+\cO\left(\frac{1}{y}\right)+\cO\left(e^{-2(x-y)}\right) \right] \ ,
\eeqn
where we have used the fact that in the large drop regime $k$ must be smaller than $k_u$, therefore $k/h$ is always small. We shall see that this transition to an ``inverse Gibbs-Thomson regime" indeed affects the system behaviour.

\subsubsection{Stability}
We expand \eqs \eqref{eq:PRL:SI:f1} for $x\gg y \gg 1$: 
\beqn
f_1 &=&- 1+\cO\left(ye^{-2(x-y)} \right) \\
f_2 &=&  1+\cO\left(\frac{1}{x}\right)+\cO\left(ye^{-2(x-y)}\right)\\
f_3 &=& -y\left(1+\cO\left(\frac{1}{y}\right) \right) \\
f_4 &=& -y \left( 1+\cO\left(\frac{1}{y}\right) +\cO\left(e^{-2(x-y)} \right) \right).
\eeqn
From the definitions of $H_{\rm in/out}$ (\eqs \eqref{eq:PRL:SI:Hin},\eqref{eq:PRL:SI:Hout}) we have
\beqn
\Ho-\Hin=-\chi\hatPin \left(1+\cO(\chi)+\cO\left(\frac{\hatPo}{\hatPin}\right) \right) \ ,
\eeqn
and by using the steady-state radius \eq \eqref{eq:PRL:SI:steadyLargeR} in the definitions of $H_{\rm in/out}$, we find
\beqn
\cO\left(\Hin\right)=\cO\left(\Ho\right)=\cO\left(\hatPin \chi\right) \ .
\eeqn
Therefore \eq \eqref{eq:PRL:SI:g1} becomes
\beqn
g_1 &=& 4 \pi D \left(-\chi \hatPin \left(1+\delta_1 \right) + \frac{2 l_c \hatPo}{\xi} (1+\delta_2) \right) \\
&=& 4 \pi D \left(-\chi \hatPin \left(1+\delta_1 \right) + \frac{2 l_c \hatPo h^{1/2}}{D^{1/2}} \left( 1+ \delta_2 \right) \right) \ ,
\eeqn
with
\beqn
\delta_1&=&\cO(\chi)+\cO\left(\frac{\hatPo}{\hatPin}\right) +  \cO\left(\frac{1}{x}\right)  +  \cO\left(ye^{-2(x-y)} \right) \\
\delta_2&=&\cO(\chi) +\cO\left(\frac{1}{y}\right) +\cO\left(e^{-2(x-y)} \right) 
\ .
\eeqn
Remembering that $\chi\equiv k/h$, we see that the system is unstable at small $k$ ($g_1>0$) and stable at large $k$ ($g_1<0$). We seek the critical rate $k_l$ at which the stability-instability transition occurs ($g_1(k_l)=0$):
\beqn
\label{eq:PRL:SI:kl}
k_l&=& \frac{2l_c \hatPo h^{3/2} }{D^{1/2}\hatPin} \left( 1+ \cO\left(\frac{k_l}{h}\right)+\cO\left(\frac{\hatPo}{\hatPin}\right) +  \cO\left(\frac{1}{y}\right)  +  \cO\left(ye^{-2(x-y)} \right) \right)
\ .
\eeqn
Interestingly the rate $k_l$ at which the system transition from the unstable to the stable regime is the same rate at which the system transition from the Gibbs-Thomson regime to the ``inverse Gibbs-Thomson regime" (\eq \eqref{eq:PRL:SI:inverseGibbs}.

If $k_l<k_u$ then $\cO(k_l/h)$ is always small (\eq \eqref{eq:PRL:SI:ku})). On the contrary when $k_l>k_u$, then $\cO(k_l/h)>1$ and $k_l$ is not defined anymore since large drops dissolve for $k>k_u$. Therefore there exists a critical backward rate $h_0$ associated to this transition, which we will now discuss.

\subsubsection{Critical backward rate $h_0$}
\label{sec:PRL:SI:h0}

We have seen that large drops ($R^*>\xi$) can exist when the forward rate $k$ is smaller than the critical rate $k_u$ (\eq \eqref{eq:PRL:SI:ku}) and are unstable to Ostwald ripening for $k<k_l$ and stable for $k>k_l$ (\eq \eqref{eq:PRL:SI:kl}). When the backward rate $h$ is larger than a critical value $h_0$, the unstability-stability transition rate $k_l$ falls outside the region of existence of the large drop regime ($k_l>k_u$), and is therefore undefined. In this case large drops are always unstable. We find $h_0$ by solving $k_l(h_0)=k_u(h_0)$:
\beqn
\label{eq:PRL:SI:hc}
h_0=\frac{D}{l_c^2} \left(\frac{\phi-\hatPo}{\hatPo}\right)^2  \left[1+\cO\left(\frac{\phi-\hatPo}{\hatPin}\right) + \cO\left(\frac{\hatPo}{\hatPin}\right)+\cO\left(\frac{1}{y}\right)+\cO\left(e^{-2(x-y)}\right) \right]\ .
\eeqn

By expressing the gradient length scale $\xi$ for $h=h_0$ we find another interesting transition associated to  $h_0$:
\beqn
\xi(h_0)&=&\sqrt{\frac{D}{k+h_0}} \\
&=& \sqrt{\frac{D}{h_0}}\left(1+\cO\left( \frac{k}{h_0} \right) \right) \\
&\simeq& \frac{\hatPo l_c}{\phi-\hatPo} \\ 
&=& R_n \ ,
\eeqn
where we used again the fact that $k/h$ is always small in the large drop regime since $k<k_u$ (\eq \eqref{eq:PRL:SI:ku}) and where $R_n$ is the radius of the nucleus in equilibrium conditions (\eq \eqref{eq:PRL:SI:Rc}), since $P_{\rm tot}\simeq \phi$ for $k<k_u$ (eq \eqref{eq:PRL:SI:Ptot}). In other words, when $h>h_0$, the gradient length scale $\xi$ is smaller than the equilibrium nucleus $R_n$.  Since in a non-equilibrium system the size of the nucleus is larger than in an equilibrium system (\sect \ref{sec:PRL:SI:noneq:intro}), the situation $h>h_0$ corresponds to the case where drops are always larger than $\xi$.

We have shown in the regime $k<k_u$ that when the backward rate is larger than the critical value $h_0$ drops are always larger than the gradient length scale $\xi$ and unstable to Ostwald ripening. We will later see that $h_0$ is also associated to a transition for small drops ($R < \xi$) in the $k>k_u$ regime.


\subsection{Small drops and low drop number density ($R\ll \xi$ and $L \gg \xi$)}
\label{sec:PRL:SI:smallDrops}
We now consider the regime where the drop radii $R$ are small and the inter-drop distance $L$ is large, compared to the gradient length scale $\xi$.

\subsubsection{Steady-state}
\label{sec:PRL:SI:smalldrop:lowrho:steady}
We expand for $x\gg1$ and $y\ll1$ the steady-state condition (\eq \eqref{eq:PRL:SI:steadyState}):
\beqn
\frac{y^2}{3}  = \lambda \left(1+ \cO\left(\frac{1}{x}\right)  +\cO(y) \right) \ ,
\eeqn
and $\lambda$ (\eq \eqref{eq:PRL:SI:lambda}):
\beqn
\label{eq:PRL:SI:lambda_large}
\lambda = \frac{\phi-\Po(1+\chi)}{\hatPin \chi}\left( 1+\cO\left(\frac{\phi-\hatPo}{\hatPin \chi}\right)+\cO\left(\frac{\hatPin}{\frac{\phi}{1+\chi}-\hatPo}\frac{R}{L^3} \right)+\cO\left(\frac{1}{\chi} \frac{R}{L^3}\right)  \right) \ .
\eeqn
From these two results we find an expression of the steady-state drop radius:
\beqn
\label{eq:PRL:SI:Ru}
R_u &=& \sqrt{\frac{3D \left(\frac{\phi}{1+\chi}-\hatPo\left(1+\frac{l_c}{R}\right) \right)}{k \hatPin} }  \\
&& \times \left(1+\cO\left(\frac{1}{x}\right)+ \cO\left(y\right) + \cO\left(\frac{\phi-\hatPo}{\hatPin\chi}\right) + \cO\left(\frac{\hatPin}{\frac{\phi}{1+\chi}-\hatPo}\frac{R^3}{L^3}\right)+\cO\left(\frac{1}{\chi}\frac{R^3}{L^3}\right) \right) \ , \nonumber
\eeqn
and if $R$ is much larger than the capillary length $l_c$ we have:
\beqn
\label{eq:PRL:SI:Ru_easy}
R_u \simeq \sqrt{\frac{3D \left(\frac{\phi}{1+\chi}-\hatPo \right)}{k \hatPin} } \ .
\eeqn

We can now check self-consistently that the ``$\cO(.)$" quantities here and in \eq \eqref{eq:PRL:SI:lambda_large} are indeed small. We start with the condition $\cO(y) \ll 1$ by comparing $R_u$ to $\xi$ which provides a lower bound on the rate $k$ for this regime:
\beqn
k &\gg& \frac{\phi-\hatPo}{\hatPin} h~ \approx k_u \ .
\eeqn
Note that $k_u$ is the upper bound on the rate $k$ for the large drop regime ($R \gg \xi$, \eq \eqref{eq:PRL:SI:ku}). This together with the fact that $\hatPin > \phi > \hatPo$ must be true in a phase-separating system shows that $\cO((\phi-\hatPo)/\hatPin\chi)$ is small. $\cO(1/x)$, $\cO[\hatPin /(\phi/(1+\chi)-\hatPo)R^3/L^3]$ and $\cO[R^3/(\chi L^3)]$ can be set arbitrary small by increasing $L$, or equivalently by decreasing the drop number density $\rho$.

\subsubsection{Critical forward rate $k_c$}

When the forward rate $k$ increases the steady-state drop radius $R_u$ decreases and falls to zero for large enough $k$. The critical rate $k_c$ at which this transition occurs can be estimated by solving $R_u(k_c)=0$ (\eq \eqref{eq:PRL:SI:Ru}).
\beqn
R_u&=&0 \\
\Rightarrow \ \ a R^3 +b R^2 + c R + d &=& 0 \ ,
\eeqn
with
\beqn
a&=&\frac{k_c \hatPin}{3 D} \\
b&=& 0 \\
c&=&- \left(\frac{\phi}{1+\frac{k_c}{h}}-\hatPo \right) \\
d&=&\hatPo l_c \ .
\eeqn
This is a cubic equation in $R$ and according to the signs of the coefficients $a,b,c,d$ there are either two real positive solutions if the determinant $\Delta =18abcd-4b^{3}d+b^{2}c^{2}-4ac^{3}-27a^{2}d^{2}$ is positive, and no real positive solutions if $\Delta<0$. We ignore complex or negative solutions since they are unphysical. The expression of $\Delta$ is:
\beqn
\Delta= \frac{k \hatPin}{D} \left[ \frac{4}{3}\left( \frac{\phi}{1+\frac{k}{h}}-\hatPo \right)^3 -   \frac{3 k \hatPin \hatPo^2 l_c^2 }{D}   \right] \ .
\eeqn
At small rate $k$ the discriminant $\Delta$ is positive so two steady-state radii $R$ exist, the larger radius being $R_u$. At large $k$ the discriminant $\Delta$ becomes negative so there are no steady-state radii and therefore no drops can exist in the system. The critical rate $k_c$ at which this transition occurs is solution of $\Delta(k_c)=0$:
\beqn
\label{eq:PRL:SI:kc:general}
\frac{k_c}{ \left( \frac{\phi}{1+\frac{k_c}{h}}-\hatPo \right)^3} = \frac{4}{9} \frac{D}{l_c^2} \frac{1}{\hatPin \hatPo^2} \ .
\eeqn
We can find upper bounds on $k_c$ by noticing the two following elements: first, this equation admits a solution only if
$\phi / ( 1+k_c/h) -\hatPo > 0$, and second, $k_c$ is a monotonic and increasing function of $h$ therefore $k_c$ is upper bounded by $k_c(h\rightarrow \infty)$. The critical rate $k_c$ is thus bounded as follow:
\beqn
\label{eq:PRL:SI:kc:bounds}
k_c < {\rm min} \left[\frac{\phi-\hatPo}{\hatPo} h \sep \frac{4}{9} \frac{D}{l_c^2} \frac{ \left(\phi-\hatPo \right)^3}{\hatPin \hatPo^2} \right] \ .
\eeqn 
In the case where $h>h_0$ (\eq \eqref{eq:PRL:SI:hc}) the critical rate $k_c$ becomes smaller than $k_u$. Since $k_c$ is only defined for small drops ($R \ll \xi$) and since only large drops exist for $k<k_u$ and $h>h_0$ (\sect \ref{sec:PRL:SI:h0}), then $k_c$ is not defined anymore in this case and drops dissolve only when $k>k_u$.

Note that we used the fact that $R\ll \xi$ in this regime to derive $k_c$. However this approximation becomes less accurate when $h$ is close to $h_0$ since the size of all drops becomes comparable to or larger than the gradient length scale $\xi$ (\sect \ref{sec:PRL:SI:h0}). Therefore when $h \approx h_0$ the critical rate $k_c$ as expressed in \eqs \eqref{eq:PRL:SI:kc:general} and \eqref{eq:PRL:SI:kc:bounds} is only a rough approximation.

\subsection{Small drops and high drop number density ($R\ll \xi$ and $L\ll\xi$)  in the $k \gg k_u$ regime}
\label{sec:PRL:SI:smallDropsHighRho}

We now study the regime where the drop radii $R$ and the inter-drop distance $L$ are both small compared to the gradient length scale $\xi$, and we moreover focus only on the $k \gg k_u$ regime (\eq \eqref{eq:PRL:SI:ku}).
\subsubsection{Steady-state}
Expanding for $y \ll x \ll 1$, the steady-state condition  \eq \eqref{eq:PRL:SI:steadyState} becomes
\beqn
\label{eq:PRL:SI:smalldrops:x1}
x^3 = \frac{(\lambda+1)y^3}{\lambda-\frac{y^2}{3}} \left( 1+ \cO\left(x^2\right)  \right) \ .
\eeqn
Imposing $y \ll x \ll 1$ on this result leads to the following requirements:
\beqn
\label{eq:PRL:SI:conditionsSmall}
&&\frac{y^2}{3} \ll \lambda \ll1  \\ 
\Rightarrow &&\cO\left(\frac{y^3}{x^3}\right) = \cO(\lambda) \\ 
\label{eq:PRL:SI:conditionsSmallBis}
\Rightarrow && x^3 \ll y
\ ,
\eeqn
and \eq \eqref{eq:PRL:SI:smalldrops:x1} thus becomes:
\beqn
\label{eq:PRL:SI:steadySmallRSmallL}
x^3=\frac{y^3}{\lambda}\left(1+\cO(x^2)+\cO\left(\frac{x^3}{y}\right)\right)
\ .
\eeqn
Plugging \eq \eqref{eq:PRL:SI:lambda} in this result we find the drop steady-state drop radius $R$:
\beqn
\label{eq:PRL:SI:smalldrop:largerho:R}
R^3&=&\frac{\frac{\phi}{1+\chi}-\hatPo}{\hatPin} L^3  
\left[1 + \frac{\chi}{1+\chi} \left( \cO\left(x^2\right) + \cO\left(\frac{x^3}{y}\right) + \cO\left(\frac{R^3}{L^3 \chi}\right) + \cO\left(\frac{\phi-\hatPo}{\hatPin \chi}\right)\right) \right. \nonumber \\ 
&& \left. +\cO\left(\frac{1}{1+\chi}\frac{\hatPo}{\hatPin} \right) \right ]  \ .
\eeqn

We now determine that the terms ``$\cO(.)$" are indeed small. $\hatPin > \phi > \hatPo$ must be true in a phase-separating system, and taking $k\gg k_u$ (\eq \eqref{eq:PRL:SI:ku}) shows that $\cO[(\phi-\hatPo)/\hatPin\chi]$ and $\cO[1/(1+\chi) \hatPo/\hatPin]$ are small. Using \eq \eqref{eq:PRL:SI:smalldrop:largerho:R} the condition $\cO(R^3/(L^3 \chi))\ll 1$ becomes:
\beqn
\chi + \chi^2 - \frac{\phi-\hatPo}{\hatPin} & \gg& 0 \\
\chi &\gg& \frac{1}{2}\left( \sqrt{1+4 \frac{\phi-\hatPo}{\hatPin}}-1\right) \simeq \frac{\phi-\hatPo}{\hatPin} \\
k &\gg& k_u \ ,
\eeqn
which is a condition we have already imposed.
Finally by using again \eq \eqref{eq:PRL:SI:smalldrop:largerho:R} the condition $\cO(x^3/y) \ll 1$ leads to a lower bound on the drop number density $\rho$:
\beqn
\rho \gg \frac{3}{4\pi}\left( \frac{k+h}{D}\right)^{3/2} \left( \frac{\hatPin}{\frac{\phi h}{k+h}-\hatPo} \right)^{1/2} \ .
\eeqn

\subsubsection{Stability}
Expanding for $y\ll x \ll1$ and keeping in mind that $x^3\ll y$ (\eq \eqref{eq:PRL:SI:conditionsSmallBis}), \eqs \eqref{eq:PRL:SI:f1} become
\beqn
f_1&=& -\frac{y^2}{3}\left(1+\cO\left(y^2\right)\right)\\
f_2&=&  \frac{x^3}{3y}\left(1+\cO(x)+\cO\left(\frac{y}{x}\right)+\cO\left(\frac{x^3}{y}\right)\right) \\
f_3&=& -\frac{y^2}{3}\left(1+\cO\left(y^2\right)\right)\\
f_4&=& -1+\cO(x)+\cO\left(\frac{y}{x}\right).
\eeqn
and using these results together with the steady-state condition \eq \eqref{eq:PRL:SI:steadySmallRSmallL}, \eq \eqref{eq:PRL:SI:g1} becomes
\beqn
g_1= 4 \pi D &&\left[-\frac{2 y^2 \Hin}{3} \left(1+\cO(x)+\cO\left(\frac{y}{x}\right)+\cO\left(\frac{x^3}{y}\right) \right) +\frac{\hatPo l_c}{R}\left(1+\cO(x)+\cO\left(\frac{y}{x}\right)\right) \right].
\eeqn
The system is unstable for small radii ($g_1>0$) and stable for large radii ($g_1<0$). The stability-instability boundary radius $R_l$ is the solution of $g_1(R_l)=0$:
\beqn
R_l y^2 \simeq \frac{3 \hatPo l_c}{2 \Hin} \ .
\eeqn
Expanding $\Hin$ gives
\beqn
\Hin=\frac{\chi \hatPin}{1+\chi} \left(1+\cO\left(\frac{\phi-\hatPo}{\chi \hatPin}\right)+ \cO\left( \frac{R^3}{L^3 \chi}\right) \right) \ .
\eeqn
and therefore we find:
\beq
\label{eq:PRL:SI:Rl}
R_l \simeq \left(\frac{3Dl_c \hatPo }{2k \hatPin} \right)^{\frac{1}{3}} 
\ .
\eeq
We check self-consistently the condition $\cO(y)\ll 1$:
\beqn
R_l &\ll & \xi \\
\frac{D^2 l_c^2 \hatPo^2}{k^2 \hatPin^2} &\ll& \frac{D^3}{h^3(1+\chi)^3} \\
k &\gg& \frac{l_c \hatPo h^{3/2}}{D^{1/2} \hatPin} \left(1+\cO\left( \frac{l_c \hatPo h^{1/2}}{D^{1/2} \hatPin}\right) \right) \simeq k_l \ .
\eeqn
When $h < h_0$ we have $k_l < k_u$ by definition (\sect \ref{sec:PRL:SI:h0}) and therefore the above condition is always true in the $k \gg k_u$ regime that we are considering in this section. Using the expression of $h_0$ (\eq \eqref{eq:PRL:SI:hc}) we also see that $\cO(l_c \hatPo h^{1/2} /(D^{1/2} \hatPin))$ is always small when $h < h_0$. If on the contrary $h>h_0$ we have already seen that only large drops exist so $R_l$ is undefined.

\subsection{Stability diagrams}
\begin{figure}
	\centering
	\includegraphics[scale=1.15]{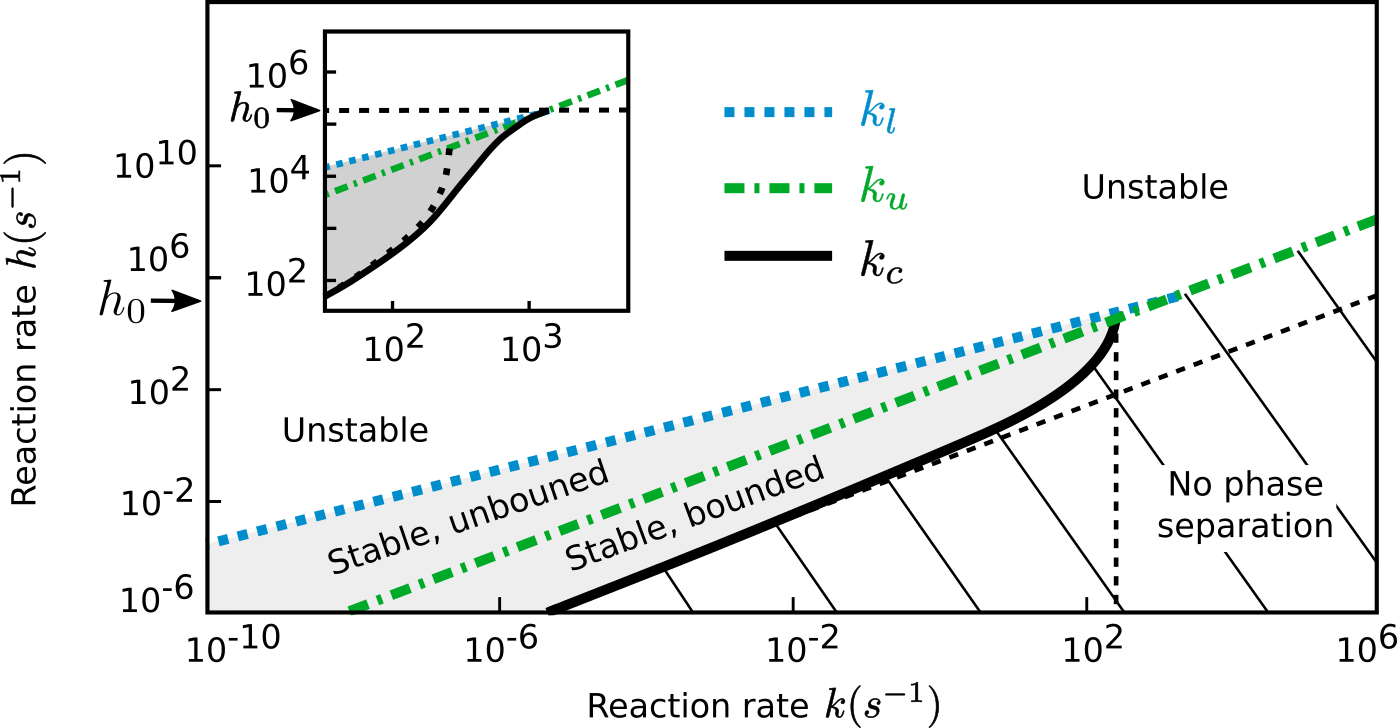}
	\caption{
		\textit{Stability diagram of a multi-drop system in the reaction rate space.} In a multi-drop system, drop existence, radius and stability, depend on the chemical reaction rates $k$ and $h$. For backward rates $h$ smaller than $h_0$ (black arrow, (\eq \eqref{eq:PRL:SI:hc})), multi-drop systems are stable against Ostwald ripening (grey region) if the forward rate $k$ is between $k_l$ (blue dotted line, \eq \eqref{eq:PRL:SI:kl}) and $k_c$ (black continuous curve, \eq \eqref{eq:PRL:SI:kc:general}). The upper bounds of $k_c$ estimated in \eq \eqref{eq:PRL:SI:kc:bounds} are shown by the black dashed lines. In the white region, multi-drop systems are unstable and coarsen via Ostwald ripening. In the stable region (grey area), the drop radius is upper-bounded if $k>k_u$ (green irregular dashed line, \eq \eqref{eq:PRL:SI:ku}), or unbounded otherwise. Phase separation is destroyed and drops dissolve in the hashed area. The validity of the expression of $k_c$ given by \eq \eqref{eq:PRL:SI:kc:general} breaks down when the drop radius $R$ approaches the gradient length-scale $\xi$. In this region, we determined $k_c$ by solving exactly the steady-state relation \eq \eqref{eq:PRL:SI:steadyState} (continuous black line in the insert figure. For comparison, $k_c$ determined from \eq \eqref{eq:PRL:SI:kc:general} is showed by the dotted black line).
		Parameters: $l_c=10^{-2}{\rm \mu m},~ D=1 {\rm \mu m^2 s^{-1}},~\phi=5.10^{-4}{ /\nu},~\hatPin=10^{-1}{ /\nu},~\hatPo=10^{-4}{/\nu}$, where $\nu$ is the molecular volume of $P$ and $S$  and can be chosen arbitrarily.
	}
	\label{fig:PRL:SI:rateDiag}
\end{figure}

We have seen that in a multi-drop system chemical reactions control drop size and coarsening, and that different regimes exist depending on the reaction rates $k$ and $h$. In \fig \ref{fig:PRL:compare} (main text) we explicit these regimes by varying the rate $k$ while $h$ is fixed. For a different choice of $h$ the system can exhibit different features as shown in \fig \ref{fig:PRL:SI:rateDiag}. 

\part{Two dimensions}
\setcounter{section}{0}
\label{sec:PRL:SI:2D}
In two dimension space ($d=2$) the steady-state condition ($g_0(R^*)=0$, \eq \eqref{eq:PRL:SI:steady:general}) becomes
\beqn
\label{eq:PRL:SI:g0:2D}
\Ho\left[A J_1(\iota y^*) - B  Y_1(-\iota y^*)\right] - \Hin \frac{ J_1(\iota y^*)}{J_0(\iota y^*)}=0
\eeqn
with $y^*\equiv R^*/\xi$, and the linearised drop growth rate of drop 1 upon perturbing the steady-state $R_1\mapsto R^*+\epsilon$, $R_2\mapsto R^*-\epsilon$ with $\epsilon \ll R^*$ is
\beqn
\label{eq:PRL:SI:g1:2D}
\frac{g_1(R^{*})}{4 \pi D R^{*2}}&=& \HHo \frac{\iota}{\xi} \left[ -A J_1(\iota y^*) + B  Y_1(-\iota y^*) \right] + \Ho \frac{\iota}{\xi} \left[ -\mathcal{A} J_1(\iota y^*) + \mathcal{B} Y_1(-\iota y^*)\right] \\
&&+ \Ho \frac{1}{\xi^2}\left[ A \left(J_0(\iota y^*)+\frac{\iota}{y^*} J_1(\iota y^*)\right) + B \left(Y_0(-\iota y^*)-\frac{\iota}{y^*}Y_1(-\iota y^*)\right)\right] \nonumber\\
&&+\HHin \frac{\iota}{\xi} \frac{J_1(\iota y^*)}{J_0(\iota y^*)}- \Hin \frac{1}{\xi^2} \left[ \frac{J_0(\iota y^*)+\frac{\iota}{y^*} J_1(\iota y^*)}{J_0(\iota y^*)}+\left( \frac{ J_1(\iota y^*)}{J_0(\iota y^*)}\right)^2\right] \nonumber
\eeqn
with $x\equiv L/\xi$ and
\beqn
A&=&\frac{Y_1(-\iota x)}{J_1(\iota x)Y_0(-\iota y^*)+ Y_1(-\iota x)J_0(\iota y^*)} \\
B&=&\frac{Y_1(\iota x)}{J_1(\iota x)Y_0(-\iota y^*)+ Y_1(-\iota x)J_0(\iota y^*)} \\
\mathcal{A}&=&\frac{\iota J_1(\iota x) Y_1(-\iota y^*)  Y_0(-\iota x) - \iota J_1(\iota y^*) Y_1(-\iota x) Y_0(-\iota x)   -\xi \frac{\HHo}{\Ho} \left[  J_0(\iota x) Y_1(-\iota x) Y_0(-\iota y^*)+  J_1(\iota x) Y_0(-\iota x) Y_0(-\iota y^*) \right]}{\xi \left[J_0(\iota x) Y_0(-\iota y^*) - Y_0(-\iota x) J_0(\iota y^*) \right] \left[J_1(\iota x) Y_0(-\iota y^*) + Y_1(-\iota x) J_0(\iota y^*)\right]} \nonumber \\
\\
\mathcal{B}&=&\frac{-\iota J_1(\iota x) Y_1(-\iota y^*) J_0(\iota x) + \iota J_1(\iota y^*)Y_1(-\iota x) J_0(\iota x)   + \xi \frac{\HHo}{\Ho} \left[ J_0(\iota x) Y_1(-\iota x) J_0(\iota y^*) +  J_1(\iota x)Y_0(-\iota x) J_0(\iota y^*)\right]}{\xi \left[J_0(\iota x) Y_0(-\iota y^*) - Y_0(-\iota x) J_0(\iota y^*) \right] \left[J_1(\iota x) Y_0(-\iota y^*) + Y_1(-\iota x) J_0(\iota y^*)\right]} \nonumber \\
\\
\Hin&=& \hatPin-\frac{ \Delta S \phi + \left(\hatPin-\Delta S\Po(R^{*}) \right)\left(1-\left(\frac{R^*}{L}\right)^2\right)}{(\chi+1)\left(1-(1-\Delta S)\left(\frac{R^*}{L}\right)^2\right)} \\
\Ho&= &\Po-\frac{\phi-\left(\hatPin-\Delta S \Po(R^{*}) \right)\left(\frac{R^*}{L}\right)^2}{(\chi+1)\left(1-(1-\Delta S)\left(\frac{R^*}{L}\right)^2\right)}.
\eeqn
Note that we are interested only in the real parts of \eqs \eqref{eq:PRL:SI:g0:2D} and \eqref{eq:PRL:SI:g1:2D}.

\part{Simulation Methods}
\setcounter{section}{0}
\section{General method}
We study the dynamics of chemically active drops in a ternary fluid using Monte-Carlo simulation methods. We consider a ternary mixture $P,S,C$ on a two-dimensional square lattice where each site has the dimension $\Delta d$. Each particle $P$ interacts with its $8$ nearest neighbours so that every $P-P$ pair contributes to the system energy by $e_{AA}$. The total system Hamiltonian is
\beqn
H=N_{PP} e_{PP}
\eeqn
where $N_{PP}$ is the total number of $P-P$ pairs in the system. We enumerate the simulation steps carried out within a simulation time unit $\Delta t$.
To simulate the system, we use the Metropolis-Hastings algorithm together with the Kawasaki exchange scheme \cite{kawasaki1972phase}. The entire lattice is searched sequentially for sites occupied by a $P$ or $S$. When such site is found, one of its $8$ nearest neighbour is randomly selected. The two sites are then exchanged with the probability
\beqn
p=
\left\{
\begin{array}{ll}
	e^{- \Delta H/(k_b T)}  \quad\quad & \Delta H > 0\\
	1  &\Delta H \le 0
\end{array}
\right.
\eeqn
where $\Delta H$ is the change in Hamiltonian caused by the exchange, $T$ the temperature and $k_b$ the Boltzmann constant.
We then consider the chemical reactions that convert $P$ into $S$ and vice versa:
\beqn
\ce{
	$P$
	<=>[k][h]
	$S$
}.
\eeqn
where $k$ and $h$ are the reaction rate constants. The entire lattice is again sequentially searched for sites occupied by a $P$ or $S$. When a site with a $P$ is found, the $P$ is destroyed and replaced by a newly created $S$, with the probability $k$. If a site with a $S$ is found, the $S$ is destroyed and replaced with a newly created $P$, with the probability $h$.

\section{Equilibrium parameters: capillary length $l_c$ and dilute phase composition $\hatPo$}
\begin{figure}
	\centering
	\includegraphics[scale=1]{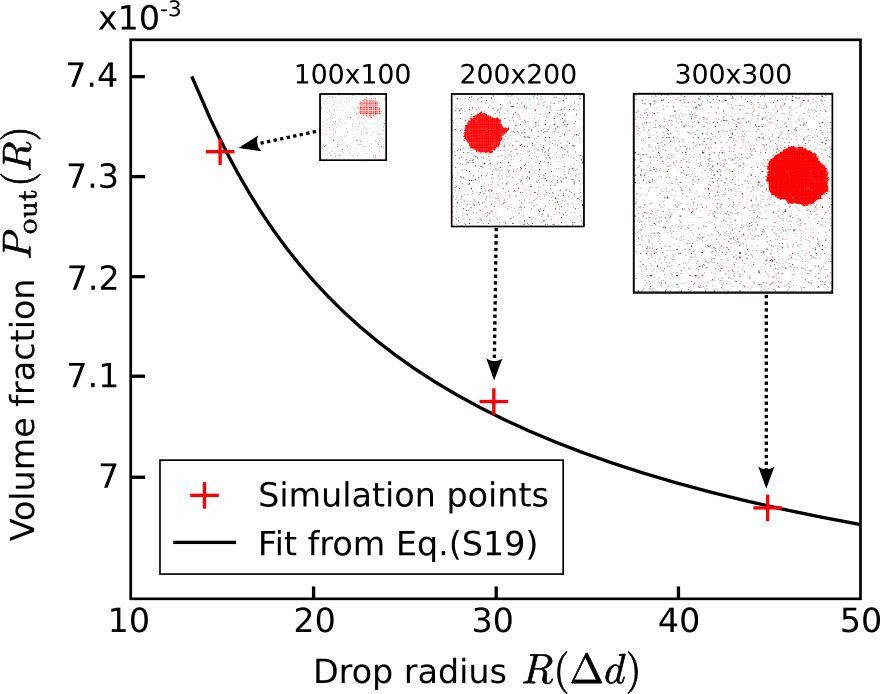}
	\caption{
		Determination of the equilibrium ($k=h=0$) parameters $l_c$ and $\hatPo$ associated to our simulations. We simulate three single-drop systems of different sizes (snapshots), extract their drop radius $R$ and concentration of $P$ in the diluted phase ($\Po(R)$), and fit the results (red points) to the Gibbs-Thomson relation (\eq \eqref{eq:PRL:SI:GibbsThomsonOut}).	 We perform a linear regression of $\Po(R)\times R = \hatPin \times R + l_c$ and find $l_c=1.2\pm 0.1 ~\Delta d$ and $\hatPo = (6.79 \pm 0.02).10^{-3}$ (see black curve for best fit). To cancel out fluctuations in $R$ and $\Po(R)$ we calculate their mean values by averaging a large number of samples. Moreover $\Po(R)$ is also spatially averaged in a square region in the dilute phase. Parameters: $P_{\rm tot}=1/13$, $S_{\rm tot}=3/130$, $e_{PP}=-9/7$, system sizes = $100 \times 100$, $200 \times 200$, $300 \times 300$ $(\Delta d)^2$.
		\label{fig:PRL:SI:capillary}	
	}
\end{figure}

To determine the equilibrium ($k=h=0$) parameters $l_c$ and $\hatPo$ associated to our simulations, we simulate single-drop systems of different sizes and extract their drop radius $R$ and $P$ concentration $\Po(R)$ in the dilute phase, and fit these results to the Gibbs-Thomson relation (\eq \eqref{eq:PRL:SI:GibbsThomsonOut}) (\fig \ref{fig:PRL:SI:capillary}). 

\section{Non-equilibrium concentration profiles}

\begin{figure}[]
	\centering
	\includegraphics{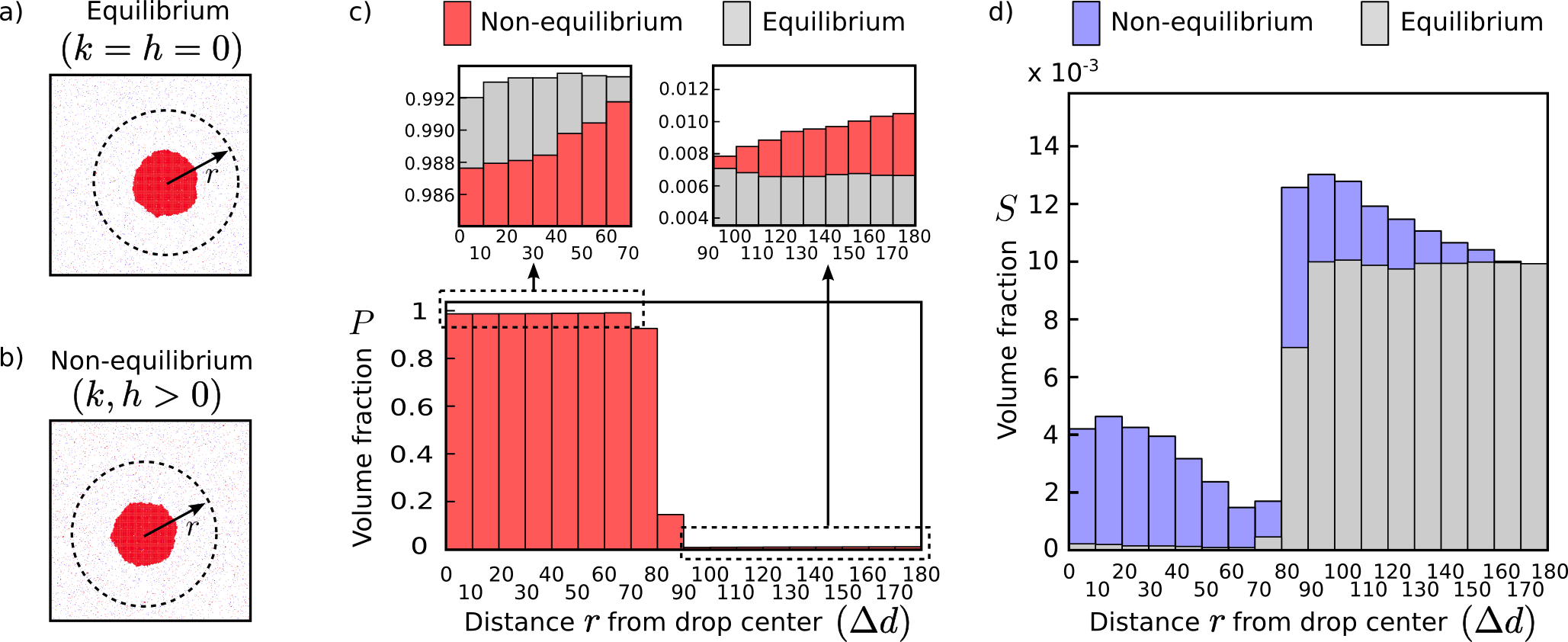}
	\caption{
		Volume fraction profiles in a single-drop system for equilibrium ($k=h=0$) (a)) and non-equilibrium ($k,h>0$) conditions (b)). c) At the drop interface, the $P$ profiles are similar both at equilibrium and non-equilibrium conditions. At non-equilibrium conditions, concentration gradients in $P$ and $S$ exist inside and outside drops (c) and d)). The profiles are radially averaged inside a disc centred on the drop center of mass (dashed line in a) and b)), then averaged over multiple samples. Parameters: system size=$500{\rm X}500~ (\Delta d)^2$, disc radius=$180 \Delta d$, $\phi=0.1$. $\epsilon_{PP}=-9/7$. Equilibrium parameters: $P_{\rm tot}=1/11$, $S_{\rm tot}=1/110$. Non-equilibrium parameters: $k=2 \times 10^{-6} (\Delta t)^{-1} $, $h=2\times 10^{-5} (\Delta t)^{-1}$.
		\label{fig:PRL:SI:profiles}
	}
\end{figure}

Our analytical work is based on the assumptions that the system is close to equilibrium and that local thermal equilibrium remains valid. The subsequent concentration profiles \eqs \eqref{eq:PRL:SI:profilePin}-\eqref{eq:PRL:SI:profilePout} obey the equilibrium conditions at the drop's interfaces (\eqs \eqref{eq:PRL:SI:GibbsThomsonIn},\eqref{eq:PRL:SI:GibbsThomsonOut}) and contain spatial gradients inside and outside drops. Seeking for validation of these assumptions we study the concentration profiles in a single-drop system for equilibrium conditions ($k=h=0$) and non-equilibrium conditions ($k,h>0$) (\fig \ref{fig:PRL:SI:profiles}). We obtain confirmation that the coexistence concentration of $P$ at the interface are similar at equilibrium and non-equilibrium, and that the profiles of $P$ and $S$ contain spatial gradients inside and outside the drop at non-equilibrium conditions.

\section{Stability-instability boundary radius}
\begin{figure}[]
	\centering
	\includegraphics[]{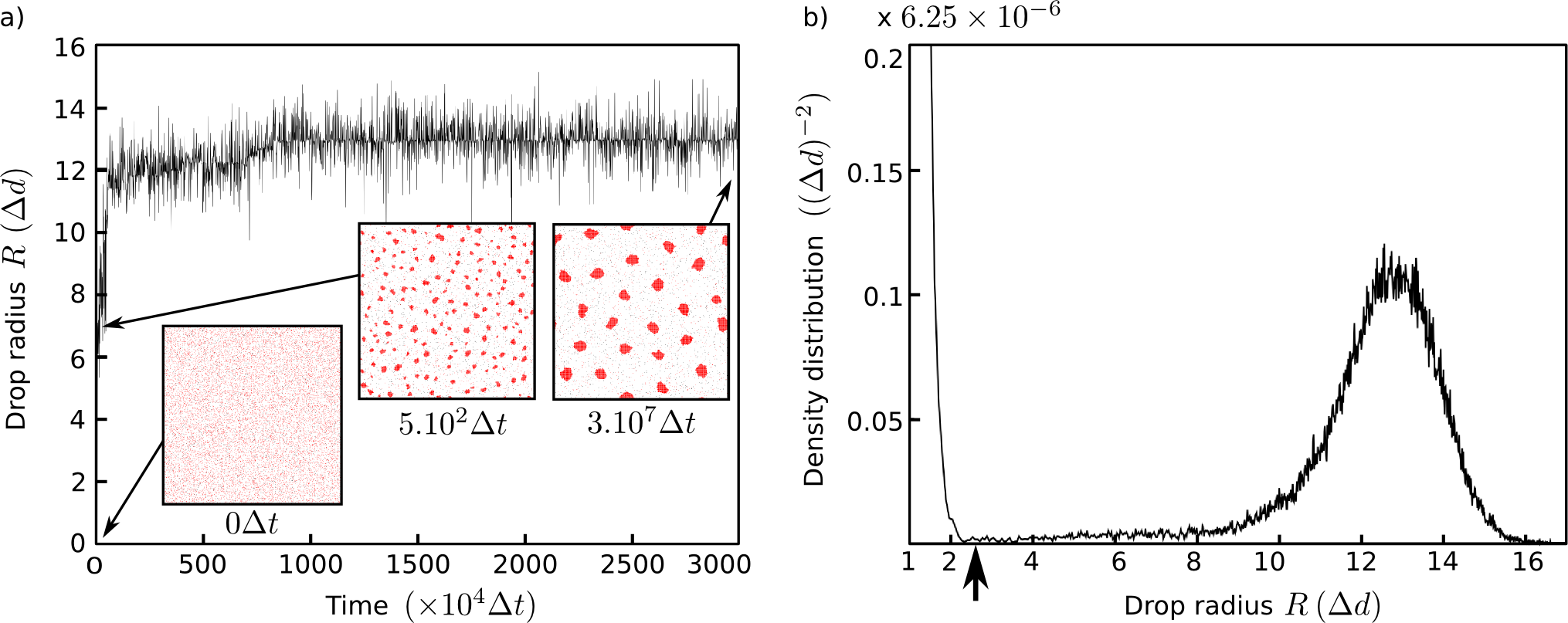}
	\caption{
		Determination of the stability-instability boundary radius. At $\Delta t=0$ particles $P$ and $S$ are randomly distributed on the lattice, ensuring global chemical equilibrium ($P_{\rm tot}=h\phi/(k+h)$, $S_{\rm tot}=k\phi/(k+h)$). a) In the early stage drops nucleate and grow, then drops undergo coarsening via coalescence and Ostwald ripening leading to an increase of the average drop radius. Eventually coarsening is arrested and the system reaches a steady-state composed of drops with similar radii. This particular steady-state radius, that is reached when starting with small drops, is defined as the stability-instability boundary radius. Snapshots (inserts) are taken at different times and $P$ and $S$ particles and shown with red dots and blue dots, respectively.  b) The steady-state radius is defined by the location of the highest peak in the drop radius distribution. The radius distribution is averaged during the second half of the simulation. We neglect the small drops that form transiently due to the stochastic fluctuations of the concentrations by ignoring drops that contain less than $20$ $P$ molecules (arrow). Parameters: system size= $400\times400 ~ (\Delta d)^2$, $\phi=0.1$, $e_{PP}=-9/7$, $h=10^{-4} (\Delta t)^{-1}$, $k=10^{-5}(\Delta t)^{-1}$.
		\label{fig:PRL:SI:radiusTimeSerie}	
	}
\end{figure}

We seek the stability-instability boundary radius (dashed line in \fig \ref{fig:PRL:compare} and \eq \eqref{eq:PRL:Rl}, main text). At time $0$ we randomly distribute $P$ and $S$ molecules on the lattice in such a way that the system is inside the phase boundary ($P_{\rm tot}>\hatPo$, \fig \ref{fig:PRL:intro}, main text) and globally at chemical equilibrium ($P_{\rm tot}=h\phi/(k+h)$, $S_{\rm tot}=k\phi/(k+h)$). In the early stage drops nucleate and grow, then drops undergo coarsening via coalescence and Ostwald ripening leading to an increase of the average drop radius. Eventually coarsening is arrested and the system reaches a steady-state composed of drops with similar radii. This particular steady-state radius, that is reached by starting from small drops, is defined as the stability-instability boundary radius. The coarsening and steady-state regimes are shown in \fig \ref{fig:PRL:SI:radiusTimeSerie}.

\section{Comparison between theoretical results and simulations}
\label{sec:PRL:SI:comparison}
We now compare our simulations to our theoretical predictions for 2D systems (Part \ref{sec:PRL:SI:2D}). Specifically we analyse the stability-instability boundary radius. We first establish the correspondence between the time and length units in the simulation ($\Delta t$, $\Delta d$) and the physical units (seconds, meters). The diffusion coefficient associated to a random walk on our lattice is given by
\beqn
D=\frac{(\Delta d)^2}{2 \Delta t}.
\eeqn
Equating $D$ to the typical protein diffusion coefficient in the cytoplasm, $1 {\rm \mu m^2 . s^{-1}}$, and $\Delta d$ to the typical protein size, $10 \rm nm$, we express the physical time and length in terms of $\Delta t$ and $\Delta d$
\beqn
1 {\rm s} &=& 2.10^4 \Delta t \\
1 {\rm \mu m}&=& 10^2 \Delta d.
\eeqn
Using this correspondence the parameters $l_c$ and $\hatPo$ (see \fig \ref{fig:PRL:SI:capillary}) become
\beqn
l_c&=&1.2\times 10^{-2} \rm \mu m \\
\hatPo&=&7.79 \times 10^{-3}\quad\\
\eeqn
Analysing the concentration profiles at the interface (\fig \ref{fig:PRL:SI:profiles}) we approximate
\beqn
\hatPin&=&1 \\
\Delta S &=&0.1
\eeqn
Using these parameters in our theoretical predictions for 2D systems (see Part \ref{sec:PRL:SI:2D})  we determine the steady-state radius $R^*$ (see \eq \eqref{eq:PRL:SI:g0:2D}) and their stability (see \eq \eqref{eq:PRL:SI:g1:2D}) as functions of the rate $k$ and for fixed rate $h$. We show the phase stability diagram in \fig \ref{fig:PRL:SI:analVsNume}, where the dashed line represents the stability-instability boundary radius. We compare this boundary to our simulation results (red points in \fig \ref{fig:PRL:SI:analVsNume}) and find a good agreement between theory and simulations. Note that we studied the regions close to the critical rates $k_c$ (\fig \ref{fig:PRL:SI:analVsNume}(a)) and $k_l$ \fig \ref{fig:PRL:SI:analVsNume}(b)) with two different choices of $h$ in order to avoid excessively large simulation times.

\begin{minipage}{\linewidth}

	\begin{figure}[H]
		\centering
		\includegraphics[]{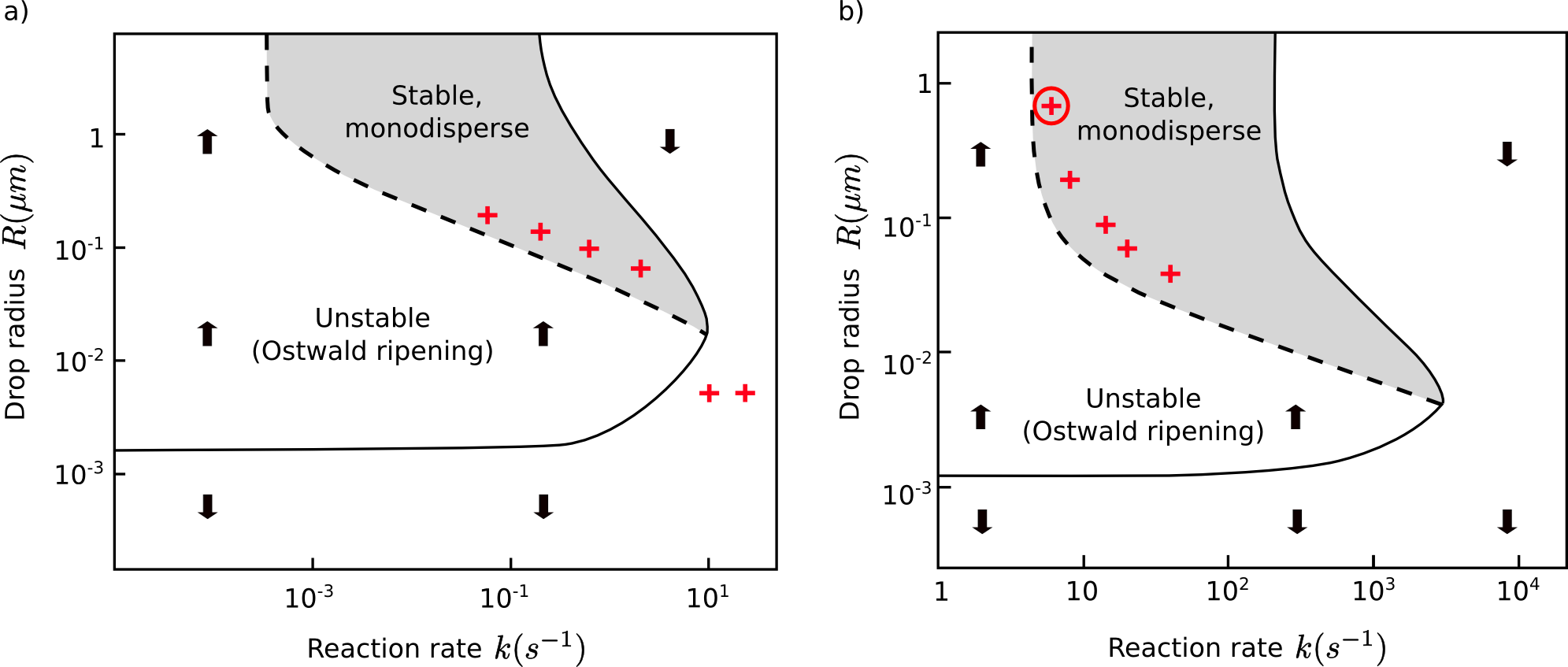}
		\caption{
			Comparison between 2D theoretical predictions and numerical simulations. The rate $k$ is varied keeping the rate $h$ fixed. A steady-state drop radius $R^*$ (solution of $g_0(R^*)=0$, \eq \eqref{eq:PRL:SI:g0:2D}) exists in the region enclosed by the continuous line. Outside this region no steady-states exist and drops dissolve (downward arrows). 
			The steady-state $R^*$ is stable inside the grey region ($g_1(R^*)<0$, \eq \eqref{eq:PRL:SI:g1:2D}). Outside the grey region the steady-state is unstable to Ostwald ripening ($g_1(R^*)>0$) causing the average drop radius to increase (upward arrows). The stability-instability boundary ($g_1(R^*)=0$) is shown with a dashed line. Regarding the simulations, the lattice is initialized at $\Delta t=0$ by randomly distributing $P$ and $S$ on the lattice in such a way that the system is inside the phase boundary ($P_{\rm tot}>\hatPo$, and see \fig \ref{fig:PRL:intro} in main text) and globally at chemical equilibrium ($P_{\rm tot}=h\phi/(k+h)$, $S_{\rm tot}=k\phi/(k+h)$). In the early stage drops nucleate, grow and coarsen, leading to an increase of the mean drop radius, then coarsening is stopped and the system reaches a steady-state defined as the stability-instability boundary (\fig \ref{fig:PRL:SI:radiusTimeSerie}). Simulation data are shown in red. The two rightmost crosses in a) represent the size of the lattice site ($\sim10^{-2} \rm{\mu m}$), i.e., there are no drops in system. The encircled cross in b) indicates that the system coarsened until a single drop remained even in the largest system simulated. There is a good agreement between theory and simulations. The duration of simulations range from $1.8\times 10^{7}\Delta t$ to $2\times 10^{8} \Delta t$. Parameters: $\phi=0.1$, $D=1\rm{\mu m^2 s^{-1}}$, $\hatPin=1$, $\hatPo=7.79\times 10^{-3}$, $\Delta S=0.1$, $l_c=1.2\times 10^{-2} \rm{\mu m}$, $e_{PP}=-9/7$. Figure a): system size=$400X400 ~({\rm\Delta d})^2 $, $h=2 ~s^{-1}$. Figure b): system size=$300 \times 300 ~({\rm\Delta d})^2$ to $400 \times 400~({\rm\Delta d})^2$, $h=2\times 10^{3} ~s^{-1}$.  See \sect \ref{sec:PRL:SI:comparison} for the equivalence between simulation and physical units.
			\label{fig:PRL:SI:analVsNume}	
		}
	\end{figure}

\end{minipage}

\end{document}